\documentclass{aa}  

\usepackage{graphicx}
\usepackage{subfigure}
\usepackage{amssymb}
\usepackage[english]{babel}
\usepackage[varg]{txfonts}
\usepackage{chngpage}
\usepackage{natbib}
\usepackage{threeparttable}
\usepackage{pgffor}
\usepackage{tikz}
\usepackage{lipsum}
\usepackage{rotating}
\usepackage{geometry}
\usepackage{lscape}
\usepackage{longtable}

\def\vycma{VY~CMa\xspace}
\def\cit6{CIT~6\xspace}

\def\nmlcyg{NML~Cyg\xspace}

\def\mucep{$\mu$~Cep\xspace}

\def\kms{km\,s$^{-1}$\xspace}
\def\msunyr{$M_{\odot}$\,yr$^{-1}$\xspace}
\def\msun{$M_{\odot}$\xspace}
\def\lsun{$L_{\odot}$\xspace}

\begin{document} 

   \title{The circumstellar environment around the extreme Galactic red supergiant NML~Cygni: Dense, dusty, and asymmetric}
   \titlerunning{Dense and dusty CSM around red supergiant \nmlcyg}
   \author{
   E. De Beck\inst{1}\fnmsep\thanks{elvire.debeck@chalmers.se}, 
   H. Andrews\inst{1}, 
   G. Quintana-Lacaci\inst{2}, 
   W. H. T. Vlemmings\inst{1}
   }
   \institute{Department of Space, Earth and Environment, Chalmers University of Technology, 412 96 Gothenburg,  Sweden 
                 \and  Department of Molecular Astrophysics, Instituto de F\'isica Fundamental (IFF-CSIC), C/ Serrano 123, 28006 Madrid, Spain
         }

   \authorrunning{E. De Beck et al.}
   \date{Received 25 March 2025; accepted 29  April 2025}

  \abstract
   {Supernova observations imply the presence of a dense and asymmetric circumstellar environment around SN Type II progenitors, whereas the mass loss from these progenitors, namely, red supergiants, is still poorly constrained. }
   {We aim to characterise the dust and gas in the circumstellar environment of the extreme Galactic red supergiant \nmlcyg in terms of mass, morphology, and kinematics. }
   {Based on interferometric observations with NOEMA at 230\,GHz, we estimated the dust masses and temperatures, and measured the extent and morphological complexity of the circumstellar environment.}
   {We detected two strong continuum components, amounting to an estimated total dust mass of $\sim2\times10^{-3}$\,\msun located out to $\sim$2000\,AU from the star, largely beyond the dust detected at optical/infrared wavelengths. The extent of the detected CO emission supports the notion that the outflow is formed by a mass-loss rate of several $10^{-4}$\,\msunyr and that it is not primarily shaped by extreme irradiation from the Cyg~OB2 cluster it has been associated with. We have detected, but not resolved, previously unseen high-velocity components close to the star. The observations reveal a very complex circumstellar morphology and we propose that some of the detected components could be the imprint of a hitherto unknown binary companion. }
   {}


   \maketitle

\section{Introduction}
Mass loss by massive stars in their red supergiant (RSG) phase is poorly understood, yet it is important to characterise this mechanism if we are to understand the late stages of stellar evolution and the behaviour and impact of their progeny, namely, core-collapse supernovae (SNe). Mass-loss rates retrieved from spectral-energy distribution (SED) modelling of RSGs in clusters cover a range $10^{-7}-10^{-5}$\,\msunyr \citep[e.g.][]{beasor2020}. In the case of some nearby extreme Galactic RSGs, mass-loss rates can reach up to several $10^{-4}$\,\msunyr \citep[e.g.][]{andrews2022,debeck2010_comdot,Shenoy2016}. Observations of \vycma, the most-studied extremely dusty Galactic RSG, further suggest a complex mass-loss history, with a combination of steady and episodic mass loss over time and outflows with different velocity fields \citep[e.g.][]{ziurys2007_vycma_complexity,quintanalacaci2023_vycma}. 

Whereas the driving mechanism behind mass loss from lower-mass evolved stars in the asymptotic giant branch (AGB) phase is broadly understood as a combination of stellar pulsations, convection, shocks high up in the stellar atmosphere, and radiation pressure on dust particles formed close to the star \citep[e.g.][]{hoefner2018}, the underlying mechanism(s) for RSG mass loss is still unknown. Suggested ingredients include, among others, turbulent convective motions and radiation driving on molecules \citep{josselin_plez_2007}, giant convection plumes lifting material beyond the stellar gravitational potential \citep[e.g.][]{lopezariste2023_mucepConvection}, and magnetic fields \citep[e.g.][]{tessore2017_magneticfields,vlemmings2017_vycma}. 

RSG mass loss forms the circumstellar environment (CSE) that Type II SN expand into. Studies have reported a requirement for dense CSE to be present close to the progenitor star to explain the SN light curve properties  \citep[e.g.][]{khazov2016_flashspectroscopy_CSM,morozova2018_SNII_denseCSM}. The dense CSE forms the basis for the build-up of a cold, dense shell where dust can efficiently form, an effect strengthened by clumpiness in the progenitor CSE \citep[e.g.][]{inserra2011}. Furthermore, it has been suggested that geometrical asymmetries found in the time-resolved observations of SN2023ixf, a Type II SN in M101, could be linked to strong asymmetries in the pre-explosion CSE \citep[e.g.][]{singh2024_SN2023ixf_asymmetry,shrestha2024_SN2023ixf_polarimetry}. The progenitor to SN2023ixf has been described as a star with properties very close to those of \vycma \citep{jencson2023}, strongly motivating further high-angular resolution observations of Galactic RSGs with known high mass-loss rates to help understand the CSE structure of progenitor stars to Type II SN.

\nmlcyg is an extreme Galactic RSG with similar stellar properties to \vycma. Molecular maser and thermal line emission profiles and dust emission observed towards \nmlcyg are suggestive of a similarly complex circumstellar appearance and mass-loss behaviour as found for \vycma \citep[e.g.][]{etoka_diamond_2004,monnier2004,schuster2006a,schuster2009,andrews2022,singh2021, singh2022_1mm}. Characterising the CSE of this member of the class of rare, extreme Galactic RSG will help us set constraints on the mass-loss mechanisms active in RSGs and set valuable constraints on the progenitors of Type II SNe. 

In this paper, we describe observations of the Galactic RSG \nmlcyg at $\lambda=1.3$\,mm (Sect.~\ref{sect:observations}) and report on the properties of its circumstellar environment. We present the continuum emission and the derived dust properties in Sect.~\ref{sect:cont} and selected molecular-line emission and the derived outflow properties in Sect~\ref{sect:line_detections}. We discuss our findings in light of observations of other well-studied RSGs and constraints on SN progenitors and pre-SN mass loss in Sect.~\ref{sect:discussion}. We present our conclusions in Sect.~\ref{sect:conclusion}. 

\section{Observations}\label{sect:observations}
Observations were gathered on 25 February and 4 \&\ 9 March 2021 using the Band 3 receiver on the NOEMA Interferometer. The observations were taken in the A configuration with 11 antennas with baselines between $19.3-751.7$\,m. The observations gathered on 25 February were discarded due to the unstable atmospheric conditions. On 4 March, some data were flagged due to tracking issues in one antenna, leading to 2.6\,h of on-source integration. On 9 March, 1.1\,h of on-source integration was achieved. Calibration was carried out with the triple stellar system MWC349 as the flux calibrator and the quasar 2010+723 as the phase calibrator. The typical absolute flux calibration uncertainty was determined to be 10\%. 

The observations were obtained with a local oscillator frequency of 226.013\,GHz resulting in measurements of the lower sideband (LSB) at $214.4-222.6$\,GHz and the upper sideband (USB) at $229.6-237.8$\,GHz. Both LSB and USB were observed at a spectral resolution of 2$\,$MHz, corresponding to a velocity resolution of ${\sim}3$\,\kms. 

The data were reduced with the use of \textsc{gildas} \citep{CLASS}, with reduction and calibration applied via \textsc{gildas/clic}. The imaging and deconvolution were carried out with \textsc{gildas/mapping}, and additional analysis was performed with \textsc{casa} Version 6.1.1.15 \citep{casa_ref}. After the initial reduction and calibration, line identification was carried out across the spectra extracted for apertures of 8\arcsec, 1\arcsec, and 0.4\arcsec around the central pixel, using the molecular databases provided by CDMS \citep{Muller2001,Muller2005} and JPL \citep{JPL}. Spectra in LSB and USB extracted towards the star and for a 1\arcsec\/ aperture are shown in Fig.~\ref{fig:nmlcyg_noema_spectra}.

When imaging, HOGBOM cleaning was applied with robust weighting. Continuum images and spectral line cubes were generated with a pixel size of 0\farcs08. A noise limit was set to 20\% of the expected sensitivity of the primary beam. Clean components were also selected with the use of the \textsc{support} procedure within \textsc{gildas/mapping} to optimise the sensitivity of the final continuum images and spectral line cubes. 

\begin{figure}[!h]
\centering
\subfigure[
Continuum emission in LSB (colour map) and USB (white contours at 3, 5, 10, 30, 50, 100, 200 $\sigma$), with two continuum components labelled as A and B and the stellar position as a black star. The synthesised beam is shown in the bottom left corner (LSB: yellow, USB: white). Residuals in USB after subtraction of the \textsc{imfit} results are shown in green contours (at $-10, -5, -3, 3, 5, 10\,\sigma$) with dashed contours for negative signal.
 \label{fig:nmlcyg_cont} ]{\includegraphics[width=.9\linewidth]{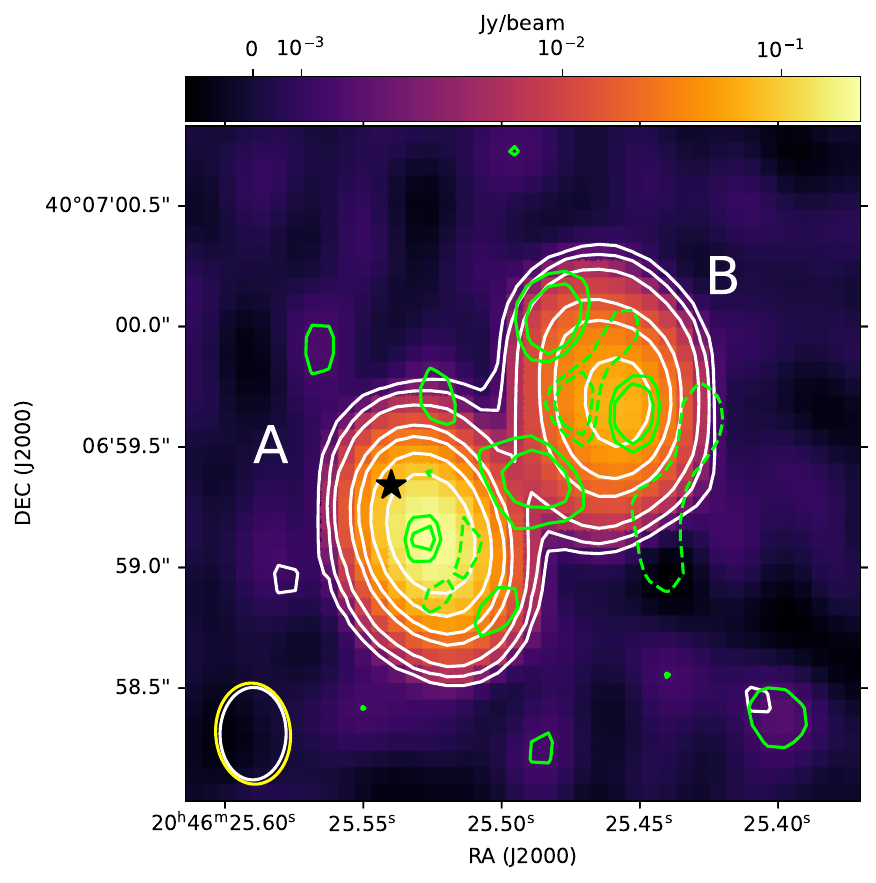}}
\subfigure[Peak flux positions of selected molecular transitions with errors according to the 50\,mas astrometric accuracy (filled circles) at systemic velocity overlaid on USB continuum contours (grey) from panel (a). Also indicated: continuum peak (purple X), stellar position extrapolated from \citet{Zhang2012} (black star), peak position of HST 0.55\,$\mu$m emission (red star), and LSB and USB beams (cyan and black ellipses). 
\label{fig:peak_pos}] {\includegraphics[width=.9\linewidth]{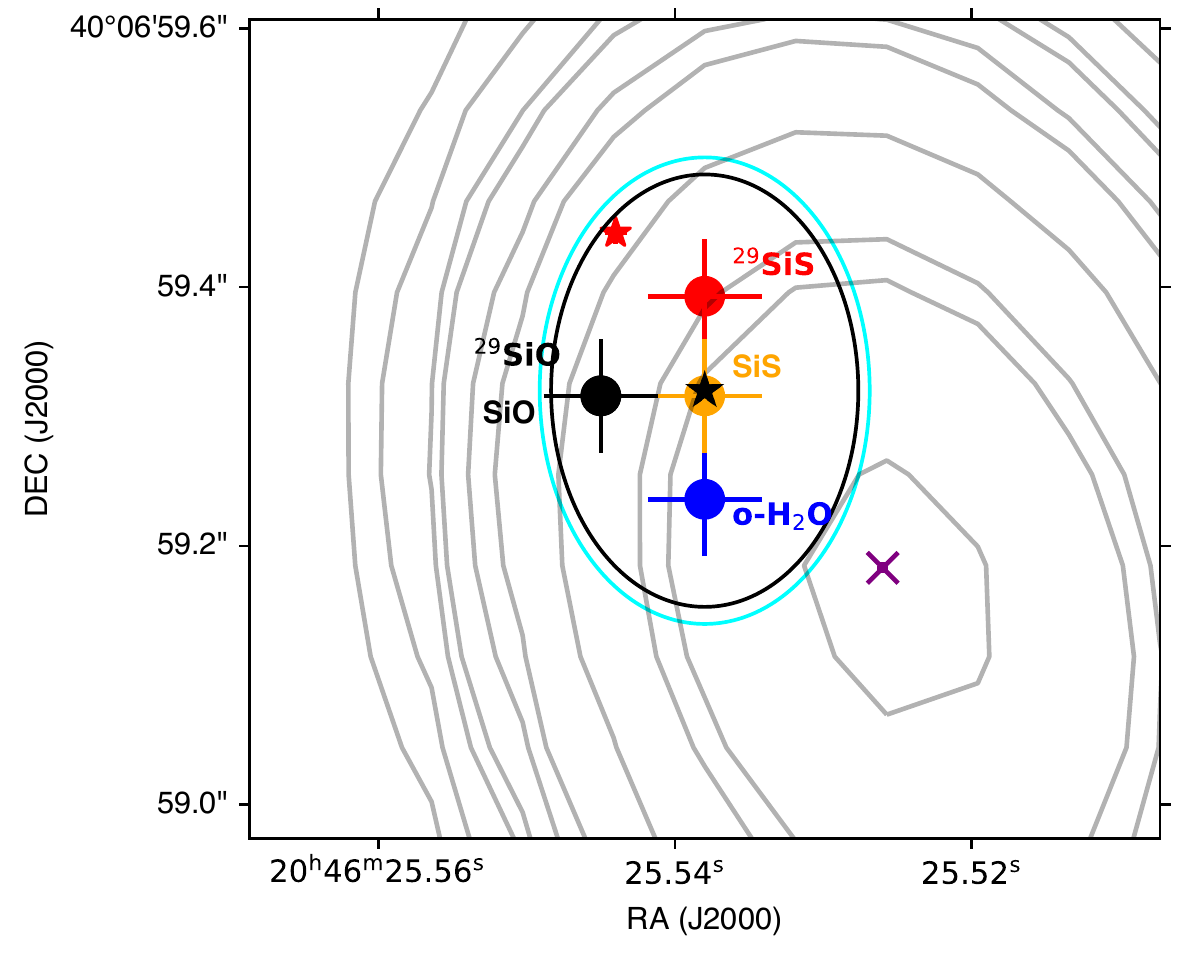}}
\caption{Continuum emission from \nmlcyg. Top: multiple components,  bottom: estimating the stellar position. \label{fig:continuum}}
\end{figure}

Final continuum images for each sideband were constructed by masking for all lines and artefacts across the total bandwidth observed.  Self-calibration solutions were applied to improve the final fidelity of the results, increasing the signal-to-noise ratio (S/N) of the continuum images by more than a factor 3. Our final continuum observations (Fig.~\ref{fig:continuum}) achieved sensitivities of $0.4-0.5$\,mJy\,beam$^{-1}$.  The synthesised beam size, noise level, and maximum recoverable scale (MRS) in the two observing bands are given in Table~\ref{tab:observations}. 

Spectral line cubes were extracted for each spectral line of interest after continuum subtraction was applied to each sideband. To provide additional information on large-scale extended structure for the CO, $^{13}$CO, SO$_{2}$, and SO lines, short-spacing observations from the IRAM 30\,m telescope using an on-the-fly map were combined with the NOEMA observations using the \textsc{gildas/uvshort} command within \textsc{gildas/mapping}. Self-calibration solutions were transferred from the continuum emission of the relevant sideband in order to improve the fidelity of the results, doubling the S/N  of the weakest line detections. The typical rms of the spectra extracted from the data with an aperture of 0\farcs4 is 2$\,$mJy at 3$\,$km$\,$s$^{-1}$ resolution and it is fairly homogeneous across the entire spectral range. 

\begin{table}[t]
\caption{Properties of the continuum observations.  \label{tab:observations} }
    \centering
    \begin{tabular}{cccccc}
        \hline\\[-2ex]
         $\nu$ & $\theta$ & $PA$ & $\sigma$ & MRS\\ 
         (GHz) & (mas\,$\times$\,mas) & ($^{\circ}$) & (mJy\,beam$^{-1}$) & (\arcsec) \\
         \hline\\[-2ex]
          218.15 & $420\times310$ & 4.5 & 0.5 & 8.85  \\
          233.75 & $384\times274$ & -0.8 & 0.5 & 8.27 \\
         \hline
    \end{tabular}
    \tablefoot{Columns list the reference frequency, $\nu$, for each sideband, the size of the synthesised beam, $\theta,$ and its position angle, $PA$, the rms noise, $\sigma$, and the maximum recoverable scale MRS of the observations.}
\end{table}

\begin{table}[t]
    \caption{Properties of continuum components A and B. }
    \label{tab:cont_prop}
    \centering
\begin{tabular}{c | cc | cc}
    \hline\\[-2ex]
                                                                                                & \multicolumn{2}{c}{Component A}                                 & \multicolumn{2}{c}{Component B}\\
                                                                                                & LSB                                                     & USB                                           & LSB                                     & USB\\
    \hline\\[-2ex]
$\theta_{\mathrm{maj}}$ (mas)   & $280.9_{1.2}$                         & $280.8_{1.7}$           & $382.3_{3.4}$    & $390.4_{5.4}$ \\
$\theta_{\mathrm{min}}$ (mas)   & $101.6_{1.7}$                         & $98.7_{2.6}$                    & $275.9_{2.2}$    & $284.7_{3.7}$ \\
$PA$ ($^{\circ}$)                               & $35.1_{0.3}$                          & $35.2_{0.5}$                    & $13.6_{1.1}$     & $12.2_{1.8}$ \\
$\Delta$ (mas)                  & $280_{30}$                     & $270_{30}$                 & $980_{30}$            & $990_{30}$ \\   
$S_{\nu}$ (mJy)                                                                         & $293.8_{0.5}$                           & $326.1_{0.9}$                 & $125.0_{0.6}$                           & $144.2_{1.2}$\\
$M_{\rm d}$ ($10^{-4}\,M_{\odot}$)& $11_{1.0}$& $11_{1.0}$& $8.2_{0.7}$& $8.9_{0.8}$\\
$T_{\rm d}$ (K) & $441_{19}$ & $448_{19}$ & $265_{11}$ & $264_{11}$\\
        \hline
        \end{tabular}
        \tablefoot{Rows list the major and minor axes, $\theta_{\rm maj}$ and $\theta_{\rm min}$, and the position angle, $PA$, for the deconvolved 2D Gaussian \textsc{imfit} components, the projected distance $\Delta$ between the component peaks and the stellar position, flux densities, $S_{\nu}$, at $\nu=218.15$\,GHz and $\nu=233.75$\,GHz,  and the estimated dust masses, $M_{\rm d}$, and dust temperatures, $T_{\rm d}$. Relevant error bars are listed in subscript.}
\end{table}

\section{Continuum emission}\label{sect:cont}
We detected two major continuum components, which we refer to as A and B in the remainder of this paper, peaking at ${\approx}475\sigma$ and ${\approx}145\sigma$, connected by a band of low-brightness emission at up to $10-20\sigma$ (Fig.~\ref{fig:continuum}), consistent between the two sidebands. Both continuum components are unresolved in the current observations and component A contains the stellar contribution. We found no evidence of any other continuum emission across the 1\farcm25\,x\,1\farcm25 field of view at the current sensitivity. We discuss the possible configuration of the system below and estimate basic properties for components A and B below.

\subsection{System configuration}\label{sect:stellarposition}
\subsubsection{Stellar position}\label{sect:stellarpos}
Using the previously determined stellar position and proper motion of the star reported by \citet{Zhang2012}, we calculated the stellar position at the epoch of the NOEMA observations to be $\alpha = 20^{\rm h}46^{\rm m}25.538^{\rm s}\pm 0.005^{\rm s}$, $\delta = 40^{\circ}06\arcmin59\farcs320 \pm 0\farcs005$. This matches the emission peaks of a selection of high-S/N and rather compact molecular emission lines from SiO, $^{29}$SiO, SiS, $^{29}$SiS, and H$_2$O well within the synthesised beam size of the observations (e.g., Figs.~\ref{fig:peak_pos} and \ref{fig:mom0_vranges_selection}) and considering the astrometric accuracy of the NOEMA observations of ${\approx}50$\,mas. We expect these emission regions to be centred on the star as, for example, predicted by \citet{menten1989_watermaser} for the H$_2$O maser at 232\,GHz and confirmed by \citet{quintanalacaci2023_vycma} for \vycma. We conclude that the expected stellar position indeed likely coincides with the actual stellar position.

We find a small offset between the peak position of the HST 550\,$\mu$m emission from \citet[][corrected for the proper motion]{schuster2006a}, and the expected stellar position. However, we note that the astrometric accuracy reported for the HST observations is of the order 0\farcs3 \citep{HST_handbook}, which does not exclude a match between this peak and the stellar position we derived. Furthermore, since those observations likely show dust-scattered stellar light, the emission peak does not necessarily coincide with the stellar position, as also discussed for the case of \vycma by \citet{Zhang2012_vycma}. 

\subsubsection{Component A}\label{sect:compA}
We used the \textsc{imfit} procedure within \textsc{casa} to retrieve the beam-deconvolved properties of the two continuum components approximated as 2D Gaussians (see Table~\ref{tab:cont_prop}). The main component, component A, is highly elongated ($\theta_{\rm min}/\theta_{\rm maj}\approx0.35$) along a roughly NE-SW axis (position angle $PA{\approx}35^{\circ}$) and encompasses the stellar position, but it peaks at a position offset from it.

Assuming the stellar parameters from \citet{Zhang2012} and assuming a spectral index of 2, the star has an expected diameter of roughly 40\,mas at the observed frequencies, that is, $10-15\%$ of the synthesised beam size, and an expected stellar flux of only ${\approx}7\%$ of the total flux of component A ($S_{\star,\rm{LSB}}=21.1$\,mJy, $S_{\star,\rm{USB}}=24.2$\,mJy). Considering this, we do not subtract the stellar contribution to recover potential substructure in component A.  

Using the stellar position described above, we find that the peak of  component A is offset by ${\approx}275$\,mas from the star (${\approx}440$\,AU at 1.6\,kpc), at a $PA{\approx}-135^{\circ}$, that is, to the SW. This offset is not resolvable by the angular resolution of the observations, but is significantly larger than both the uncertainty on the stellar position \citep{Zhang2012} and the astrometric accuracy of the NOEMA observations (${\approx}50$\,mas). We therefore suggest the presence of a bright dust clump physically offset from the stellar position in a SW direction. 

Considering the similarity in size and orientation between our component A and the eastern component measured in the mid-infrared (MIR) by \citet{schuster2009}, we propose that these measurements might be tracing the same material. This is supported by the close match between our estimated dust properties (Sect.~\ref{sect:dust}) and the results of \citet{schuster2009}. Given the considerations about the stellar position discussed above, our observations of continuum and thermal spectral line emission at 230\,GHz thus imply that the assumption by \citet{schuster2009}, namely, of the star being located at their detected MIR peak brightness, is likely incorrect. This was also suggested by \citet{Zhang2012} based on continuum and SiO and H$_2$O maser measurements at 43\,GHz and 22\,GHz. We discuss the match between our observations and the maser observations in Sect.~\ref{sect:line_detections}. 

\begin{figure*}
\centering
\subfigure[Stellar position.]{\includegraphics[width=0.3\linewidth]{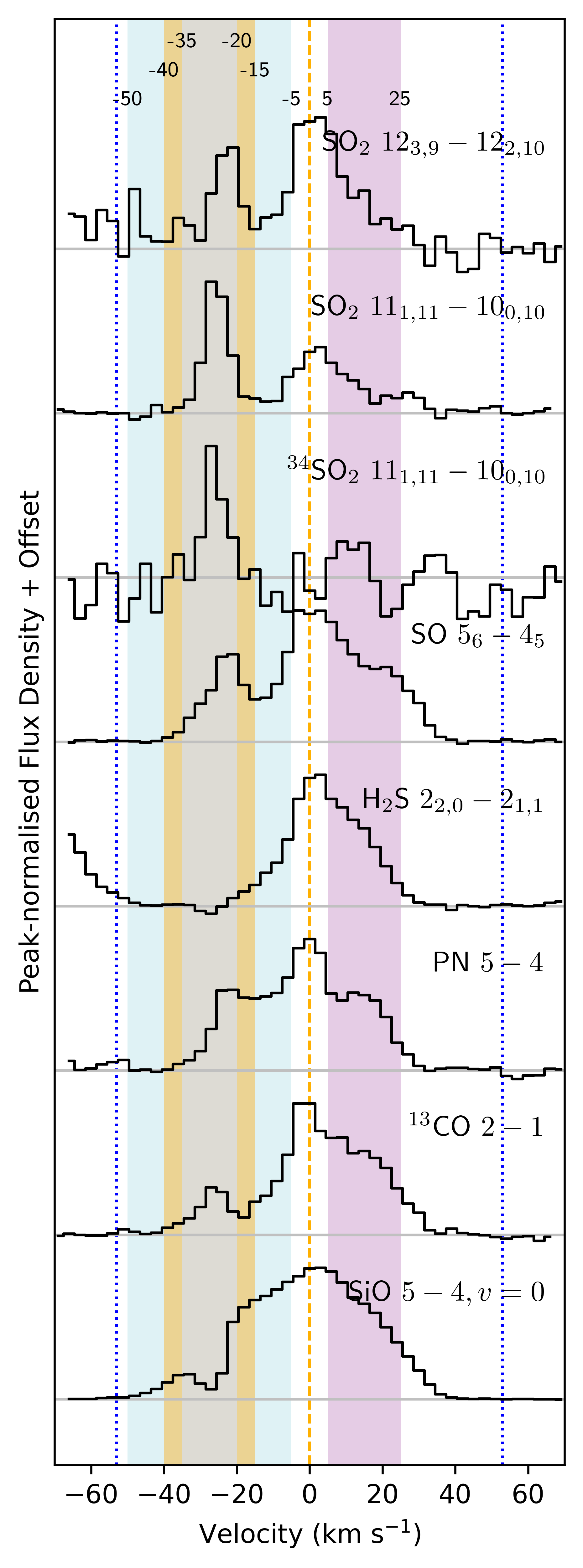}}
\subfigure[Component A.]{\includegraphics[width=0.3\linewidth]{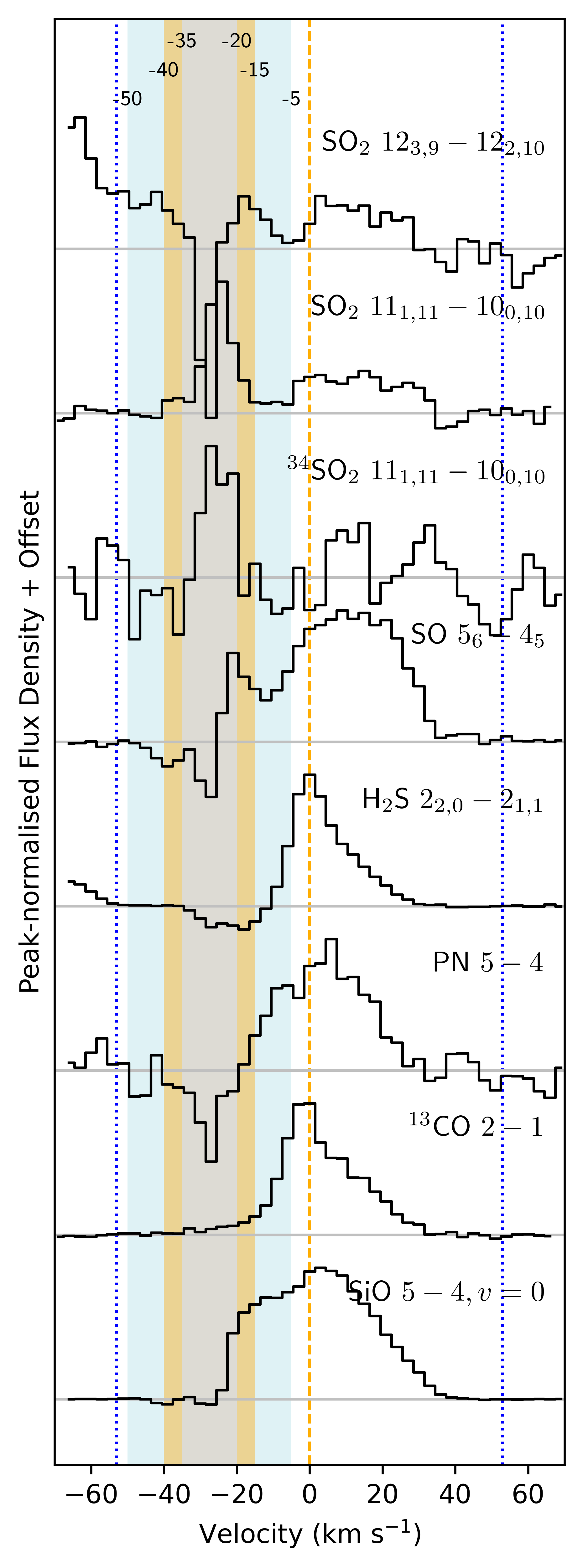}}
\subfigure[Component B.]{\includegraphics[width=0.3\linewidth]{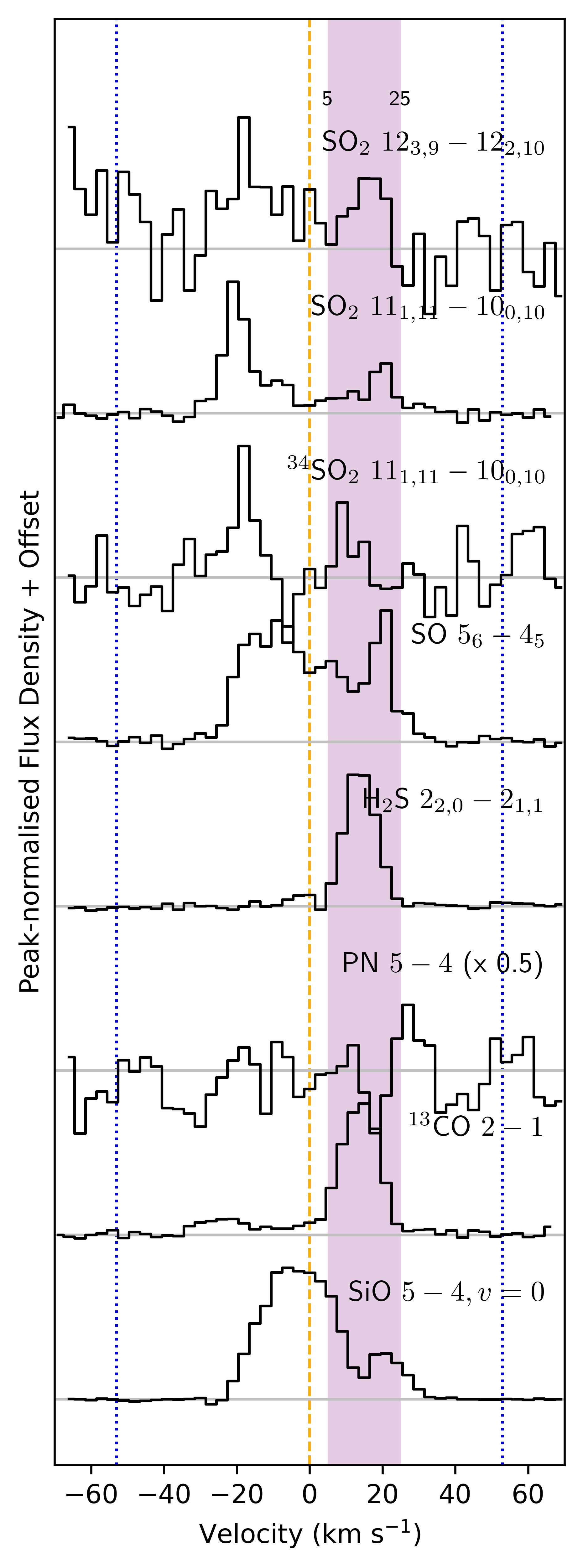}}
    \caption{Pencil-beam spectra towards NML Cyg, extracted towards the stellar position and towards the peak positions of components A and B. The spectra have been normalised to their peak and a vertical offset has been added. The transitions are labelled in each panel. Note: the spectrum of PN $5-4$ towards component B has been scaled for the sake of visualisation. Shaded regions indicate velocity ranges of interest with respect to absorption and emission features, with the relevant boundaries indicated in the top of each panel. \label{fig:extracted_spectra_absorption} }
\end{figure*}

We can use the molecular emission observed along the line of sight towards component A to constrain this component's location with respect to the plane of the sky. Pencil-beam spectra of selected transitions of SO$_2$, SO, H$_2$S, and PN, extracted towards the peak of component A reveal strong emission lines with a lowered emission and even absorption at blue-shifted velocities roughly in the range $[-40,-16]$\,km\,s$^{-1}$ (Fig.~\ref{fig:extracted_spectra_absorption}), and very clearly visible in the channel maps of CO, SiO, SO, and SO$_2$ at velocities $\lesssim-22$\,km\,s$^{-1}$ in Figs.~\ref{fig:12co-channels}, \ref{fig:SiOv0-channels}, \ref{fig:SO-5645-channels}, and \ref{fig:so2_11-10-channels}. We note here that not all of the detected SO$_2$ lines show absorption in this pencil beam towards the peak of A and we discuss this further in Sect.~\ref{sect:molecular_continuum_emission}. The absorption features can be explained by the presence of a warm body of dust behind colder gas. Although the velocity ranges covered by the absorption components are not identical across all molecular transitions, we propose that component A is located in front of the plane of the sky. The presence of emission at other (less blue-shifted) velocities along this line of sight suggests that the dust could be, at least partially, optically thin at the observed wavelengths. However, since the deconvolved size of component A is smaller than the synthesised beam, we would also detect emission originating to the sides of the component in the same beam, even if the dust were optically thick. We note that the position-velocity diagrams for a cut along $PA=65^{\circ}$ (discussed in Sect.~\ref{sect:line_detections}) show a clear indentation for the SiO $v=0$ and H$_2$S contours along the negative velocity axis at an offset of $-0\farcs3. $  This corresponds to the location of component A and suggests that the dust could indeed be optically thick at these wavelengths, effectively blocking the emission coming from a very small area on the sky. Since we do not see a similar behaviour for the SiO $v=1$ emission, which is likely more compact than the SiO $v=0$ emission, we suggest that component A is located beyond the SiO $v=1$ emission volume, relative to the star. Observations at a higher angular resolution and at significantly different wavelengths are needed to obtain a spectral index for component A to set stronger constraints on the dust properties.

\subsubsection{Component B}\label{sect:compB}
The secondary continuum component, component B, is roughly $40-45$\% as bright as and somewhat larger than component A. The \textsc{imfit} results suggest a slight elongation along a roughly NNE-SSW direction ($PA{\approx}13^{\circ}$) and an offset with respect to the stellar position of ${\approx}985$\,mas (${\approx}1570$\,AU) at  $PA{\approx}-65^{\circ}$ (i.e., to the NW of the star).  

For comparison, \citet{schuster2009} also found an elongated structure in the 8.8\,$\mu$m MIR observations, with emission from warm dust ($200-500$\,K) reaching up to $300-500$\,mas to the NW of the star, placing this warm dust between the star and our component B, where our data reveal a band of continuum emission connecting components A and B. 

We found emission from several molecular species roughly coincident with the projected location of the peak emission of component B (CO, SO, and H$_2$S; see Sect.~\ref{sect:line_detections}), with a clear prevalence of red-shifted emission, mostly in the range $[17, 23]$\,km\,s$^{-1}$. The morphology of the emission from H$_2$S and H$_2^{33}$S traces the location of component A, the `bridge' from A to B, and, finally, the location of component B when moving from blue-shifted to red-shifted velocities. A similar co-location of H$_2$S and continuum emission has been reported in observations of \vycma \citep{quintanalacaci2023_vycma}. Hence, we propose a location for component B and possibly also the dust excess measured by \citet{schuster2009},  behind the plane of the sky. As opposed to our results for component A, we found no absorption in the spectra extracted towards component B. 

\subsubsection{Geometry}\label{sect:continuumcomplexity}
Multiple mass ejections and/or dust shells have been invoked to explain a number of previous observations of \nmlcyg \citep[e.g.][]{Monnier1997,blocker_2001}. The presence of multiple continuum components was also suggested by \citet{schuster2006a,schuster2009} based on the irregular morphology found in the NIR/MIR observations and is now solidified by our mm-wavelength observations.  We note that component B reaches far beyond the 0\farcs1 dust destruction radius suggested by \citet{schuster2006a,schuster2009}. 

After subtraction of the two 2D Gaussian components, the \textsc{imfit} procedure reveals residual emission (Fig.~\ref{fig:nmlcyg_cont}) close to the peak positions of components A and B, at the NE tip of component B, and in the region connecting components A and B. Several negative residual components are also visible. Clearly, the approximation of the observations as a combination of two 2D Gaussian components is insufficient. Therefore, we consider it likely that higher-angular resolution observations will reveal substructure within the here presented continuum emission, similar to what was found for VY CMa \citep{ogorman2015, kaminski_vycma_2019,asaki2020_alma_longbaseline}. 

Components A and B are located along a NW-SE axis and are, at the same time, both elongated along a roughly perpendicular direction (NE-SW). We note that OH, H$_2$O, and SiO maser observations have revealed a similar NW-SE geometrical axis, leading to the suggestions of a possible bipolar outflow from \nmlcyg, with a receding NW lobe and an approaching SE lobe, and a rotating SiO maser shell \citep{richards_1996_masers_h2o,etoka_diamond_2004,boboltz_2000_siomasers_nmlcyg}. This orientation is in line with our suggestion that component B is located behind the plane of the sky. We refer to Sect.~\ref{sect:line_detections} for a further discussion on the envelope morphology based on our NOEMA observations and the high-angular resolution maser results from the literature.

\subsection{Dust temperature and mass}\label{sect:dust}
As argued above, dust is the dominant contributor to the observed continuum emission. To estimate the dust mass responsible for the observed flux densities, $S_{\nu}$, we assume  the dust is optically thin and that the Rayleigh-Jeans approximation is valid to obtain the dust mass, $M_{\rm d}$, from 
\begin{equation}
    S_{\nu} = \frac{3M_{\rm d}Q_{\nu}T_{\rm d}k\nu^{2}}{2a_{\rm g}\rho_{\rm g}c^{2}d^{2}},
    \label{eq:dust_mass}
\end{equation}
where $Q_{\nu}$ is the grain emissivity, $T_{\rm d}$ is the dust temperature, $k$ is the Boltzmann constant, $a_{\rm g}$ and $\rho_{\rm g}$ are the radius and mass density of the dust grains, respectively, $c$ is the speed of light, and $d$ is the distance to the star \citep{Knapp1993_dust}. In the case of component A, $S_{\nu}$ is the measured flux density of the component corrected for the expected stellar flux, $S_{\star,\nu}$ (see Sect.~\ref{sect:compA}).

We assume that the grain emissivity behaves as $Q_{\nu}  \propto \nu^{\beta}$ and $S_{\nu}\propto \nu^{\alpha}$ with $\alpha = 2 + \beta$. For the dust around \nmlcyg, we expect $\beta$-values in line with those found for other evolved stars. Previous models of the dust around \vycma have used typical values of $\beta  = 0.7-0.9$ \citep{Knapp1993_dust, ogorman2015, kaminski_vycma_2019}. Our continuum detections and the non-detection of dust at 43\,GHz by \citet{Zhang2012} are indeed in line with these literature values. We proceed to derive dust properties under the assumption that $\beta=0.9$.

Optically thin dust at a radius $r_{\rm d}$, heated by the incident stellar radiation field yields a dust temperature 
\begin{equation}
    T_{\rm d} = \left(\frac{L_{\star}T_{\star}^{\beta}}{16\pi \sigma r_{\rm d}^{2}}\right)^{1/(4+\beta)}.
\end{equation}
Assuming a stellar luminosity of $L_{\star}\approx2.7\times10^{5}\,L_{\odot}$, stellar temperature  $T_{\star}=2500$\,K \citep{Zhang2012}, and distances, $r_{\rm d}$, corresponding to the offsets $\Delta$ listed in Table~\ref{tab:cont_prop}, we derived $T_{\rm d}\approx445$\,K  for component A and $T_{\rm d}\approx265$\,K for component B. Assuming typical dust grain properties for M-type evolved stars \citep[$a_{\rm g} = 0.2\,\mu$m, $\rho_{\rm g} = 3.5$\,g\,cm$^{-3}$;][]{Knapp1993_dust}, we derived dust masses $M_{\rm d} \approx11 \times10^{-4}\,M_{\odot}$ for component A and $M_{\rm d} =8-9\times10^{-4}\,M_{\odot}$ for component B.

The assumed distances $r_{\rm d}$ are lower limits, meaning that the calculated dust temperatures, $T_{\rm d}$, are upper limits and the calculated masses, $M_{\rm d}$, are lower limits (under the assumptions on optical depth and dust emissivity). An angle to the plane of the sky of $45^{\circ}$ or $80^{\circ}$ would lead to larger spatial offsets from the star and a consequent ${\sim}15\%$ or ${\sim}50\%$ decrease in dust temperature, respectively. The derived masses, in turn, would increase by ${\sim}10\%$ and a factor of ${\sim}2$, respectively. Unfortunately, we cannot, currently, constrain these inclination angles; we refrained from deriving an inclination angle based on an assumed expansion velocity considering the outflow complexity (see Sect.~\ref{sect:line_detections}) and that we do not have information on the clump geometry.

\section{The molecular outflow}\label{sect:line_detections}
We detected 72 spectral features of which we identified 67 to pertain to 26 different isotopologues. Here, KCl and its isotopologue K$^{37}$Cl are have been identified for the first time around an M-type (carbon-poor) supergiant. An overview of the spectra is given in Fig.~\ref{fig:nmlcyg_noema_spectra}.

In this paper, we use the first spatially resolved observations of thermal emission from CO, SiO, SO$_{2}$, SO, and H$_2$S to trace the extent, morphology, and kinematics of \nmlcyg's outflow. We investigate their relation to the NOEMA continuum measurements presented in Sect.~\ref{sect:dust} and literature results, in particular, from maser emission. An in-depth presentation of the additional molecular features in this data set will be made in upcoming publications.

\subsection{Extent}\label{sect:extent}
We detect emission from $^{12}$CO and $^{13}$CO and for the first time also from C$^{18}$O towards \nmlcyg. Figures~\ref{fig:swblob}, \ref{fig:pvdiagrams}, \ref{fig:12co-channels}, and \ref{fig:13co-channels} show channel maps and position-velocity diagrams of the $^{12}$CO and $^{13}$CO($2-1$) emission that we base the rest of the discussion on.

The CSE shows significant asymmetry, with the largest CO extent measured to the SE, the smallest to the NW.  We detect $^{12}$CO($2-1$) emission at $>10\sigma$ out to 8\farcs6 from the stellar position in the SW direction, 6\farcs8 in the NE direction, 6\farcs1 in the NW direction, and 11\farcs4 in the SE direction (Fig.~\ref{fig:pvdiagrams}). This largest extent in the SE direction is also traced by a $3\sigma$ component in the $^{13}$CO emission. We note that the PV-diagrams in Fig.~\ref{fig:pvdiagrams} show $^{12}$CO emission $>3\sigma$ beyond 10\arcsec\/ along all four of these directions (SW, SE, NE, NW), but we do not consider these here, as there are artefacts in the images at similar levels (closer to the edge of the FoV) and because we cannot easily disentangle the CSE emission from the ISM contribution in the channels at $5-8$\,km\,s$^{-1}$ (Fig.~\ref{fig:co_fullfov}), where the largest CO extent is seen in Fig.~\ref{fig:pvdiagrams}. 

The detected CO angular extents of 11\farcs4 and 6\farcs1 translate to linear, radial (projected) distances from the star of $2.7\times 10^{17}$\,cm and $1.5\times 10^{17}$\,cm, respectively \citep[at 1.6\,kpc;][]{Zhang2012}. We expect these scales to be somewhat smaller than the CO photodissociation radius, by up to a factor of a few, as a consequence of the excitation and detectability of CO($2-1$), as also described by \citet{ramstedt2020_deathstar} for the densest AGB outflows. Assuming the CO abundance typically assumed for an M-type AGB star (CO/H$_2=2\times10^{-4}$) and extrapolating the results of \citet[][see their Table~B.1]{saberi2019} to a mass-loss rate of $5\times10^{-4}$\,\msunyr \citep[which would be an upper limit for \nmlcyg;][]{gordon2018_nmlcyg_dust} leads to a half-abundance radius $r_{1/2}\approx10^{18}$\,cm. This would be at the high end of the range we would find likely to be the actual photodissociation radius for \nmlcyg's outflow. Considering the proximity to Cyg~OB2, \nmlcyg plausibly resides in a more intense radiation field, which would cause a smaller molecular outflow, but likely by less than 15\% because of the high mass-loss rate \citep{saberi2019}. Furthermore, the smaller extent to the NW could possibly be due to irradiation from Cyg~OB2, although the observed SE-NW axis ($PA\approx-35^{\circ}$) is not quite aligned with the direction to the cluster ($PA\approx-63^{\circ}$). We conclude that the CO extent around \nmlcyg fits the expectations rather well for an assumed mass-loss rate and outflow velocity and that the molecular CSE extends significantly beyond the limits suggested by \citet{schuster2006a}. We also note that according to \citet{dharmawardena_2018_jcmtdust}, the transition from the stellar to the interstellar radiation field dominating grain heating occurs at $\approx0.19\mathrm{\,pc}\approx24\arcsec\/$ in the case of \nmlcyg, that is, well beyond where we detect CO($2-1$) emission.

We note that asymmetries, with a predominant SE-NW orientation, are also seen in the red-shifted SO and SO$_2$ emission. These show a clear extension to the SE, most clearly visible in the 14\,km\,s$^{-1}$ channels (Figs.~\ref{fig:so2_11-10-channels} and \ref{fig:SO-5645-channels}). An extension along a similar SE-NW axis -- but on different spatial scales -- is traced by SiO, OH, and H$_2$O maser emission \citep[$PA\approx140^{\circ}-150^{\circ}$;][]{boboltz_2000_siomasers_nmlcyg,richards_1996_masers_h2o,etoka_diamond_2004} and dust emission \citep{blocker_2001}. The observed asymmetry could be a consequence of extrinsic factors, for example, a less intense UV irradiation of the molecular gas in the SE compared to the NW direction, connected to the location of \nmlcyg relative to the Cyg OB2 cluster, and/or it could be  a consequence of factors intrinsic to the \nmlcyg system, such as a highly irregular mass loss. 

Although we refrain from discussing in detail the chemical networks in the CSE of \nmlcyg in this paper, we wish to highlight some findings from a first examination of the extent of the molecules other than CO presented in this paper. We find that the H$_2$S emitting region is strongly correlated with the continuum emission presented in Sect.~\ref{sect:cont} and only extends somewhat beyond that, out to a distance of $\approx1\farcs4$ from the star, which is significantly smaller than the $1000\,R_{*}\approx9\arcsec$ suggested by \citet{singh2022_1mm}.  Their modelling efforts, based on single-dish observations, led to the suggestion that CO, HCN, and H$_2$S have the most extended molecular abundance distributions. Furthermore, while the abundance profiles modelled by \citet{singh2021,singh2022_1mm} suggest an SiO envelope that is similar to or smaller than those observed for SO and SO$_2$, we find that the emitting region of SiO ($v=0$) reaches angular distances from the star of $2\farcs8-4\farcs5$ and is slightly larger and more circularly symmetric than the overall SO and SO$_2$ emitting regions, which exhibit more substructure along the SE-NW axis and extend out to $\sim$3\arcsec. This direction-dependent difference in extent is best seen when comparing Figs.~\ref{fig:pv_co_13co_so2_65} and \ref{fig:pv_co_sio_h2s_65}.

\subsection{Morphology}\label{sect:morphology}
The CO emission traces a complex outflow morphology, as shown in the channel maps and position-velocity diagrams in Figs.~\ref{fig:co_fullfov}--\ref{fig:13co-channels}, and \ref{fig:pvdiagrams}. We find significant substructure, visible as arcs and blobs, also in the emission of SiO, SO, SO$_2$, and H$_2$S (Figs.~\ref{fig:SiOv0-channels} -- \ref{fig:mom0_vranges_selection}). 

\subsubsection{Molecular emission co-spatial with the continuum emission}\label{sect:molecular_continuum_emission} 

We find a lack of emission and even find absorption features in CO, SiO, SO, SO$_2$, H$_2$S, and PN at blue-shifted velocities $\lesssim 16$\,km\,s$^{-1}$ in the line of sight towards the peak of component A and rather see a ring of emission surrounding this spatial region (see spectra in Fig.~\ref{fig:extracted_spectra_absorption} and channel maps in App.~\ref{sect:spectrum}). Considering that we can see a similar effect for a variety of molecules, but not all detected lines, and that it is limited to the blue-shifted part of the spectrum, we suggest that this is the result of absorption in the (colder) gas against a background of hot dust (see also Sect.~\ref{sect:compA}), rather than optically thick dust obscuring gas located behind it. Another possibility is that the affected lines are optically thick in these velocity ranges, obscuring the underlying dust emission, and that continuum subtraction (performed across all channels) causes an erroneous absorption feature, as suggested by \citet{humphreys2024_vycma_ALMA} in the case of some molecular lines observed towards \vycma. However, since we do not attempt to constrain detailed properties of the dust or gas based on these features and since component A is smaller than our synthesised beam and possibly made up of multiple smaller components, we did not explore this option further. 

We observe that the detected SO$_2$ emission shows a diverse range of behaviour: some transitions show the absorption feature towards the peak of component A mentioned above, whereas others show distinct emission in this same line of sight and velocity range (Fig.~\ref{fig:extracted_spectra_absorption}). This suggests that the excitation of SO$_2$ could be very strongly coupled to the radiation field in line with the findings of \citet{danilovich2016_sulphur}, emphasising the importance of considering the possible complexity of excitation when interpreting molecular emission maps of only selected transitions. \citet{cernicharo2011_so2} predicted and confirmed absorption and maser behaviour for a number of long-wavelength rotational transitions of SO$_2$ towards cold ($<40$\,K) molecular clouds.

In contrast to what we find for component A, the spectra towards the peak of component B do not exhibit strong absorption features. We find that the SiO emission shows a depression in a red-shifted velocity range around 20\,\kms where several other molecular transitions, including $^{13}$CO, H$_2$S, SO, and SO$_2$ conversely show rather distinct emission components. This could be suggestive of a local depletion of SiO.  

We find that the $^{12}$CO, $^{13}$CO, and H$_2$S emission trace a clear `bridge' between components A and B across the velocity range $[-7,23]$\,\kms, coincident with a `bridge' of significant continuum emission (Figs.~\ref{fig:continuum}, \ref{fig:12co-channels}, \ref{fig:13co-channels}, \ref{fig:h2s-channels}). The velocities at which we see this emission structure suggest that the connection from the stellar position to B extends behind the plane of the sky. At the same time, the SiO, SO$_2$, and SO emission do not trace this `bridge'. The SiO emission is mostly limited to the NE and SE quadrants of the maps, whereas the SO and SO$_2$ emission regions do reach significantly beyond the location of component B in the NW quadrant across the velocity range of $[-14,13]$\,\kms. The SO emission maps, however, show a distinct gap at the location of the `bridge' at redshifted velocities (most clearly seen for the $5_6-4_5$ transition at $[8,17]$\,\kms; Fig.~\ref{fig:SO-5645-channels}).

The morphology of the detected H$_2$S emission is closely related to that of the continuum, with emission components clearly spatially correlated with components A and B across velocity space and extending only little beyond these (Fig.~\ref{fig:h2s-channels}). This is in line with the finding for \vycma that the presence of H$_2$S appears strongly connected to the presence of warm dust \citep{quintanalacaci2023_vycma}. Higher-angular resolution observations are needed to resolve the structure of both the gaseous and dusty component to establish how close to the star H$_2$S actually is formed and how this relates to the presence of dust. In the case of AGB stars, H$_2$S is thought to be formed at high densities in the inner layers of the outflows of oxygen-rich stars with high mass-loss rates \citep{danilovich2017_h2s}. According to the models by \citet{vandesande2022_companionUV}, the presence of H$_2$S and a shell-like distribution of SO, possibly quite similar to what we detected for \nmlcyg, rule out a very strong UV field from a companion star. 

\begin{figure*}[!ht]
\centering
    \includegraphics[width=\linewidth]{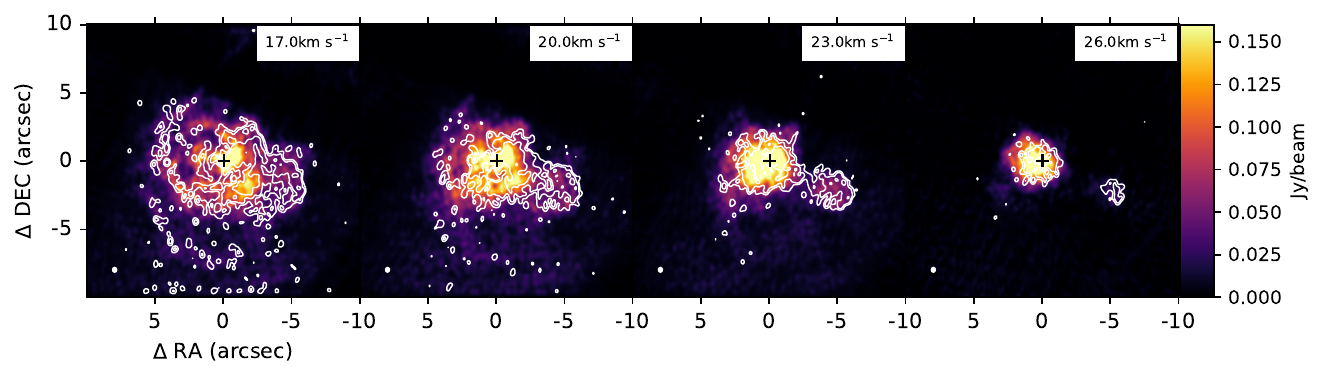}
    \caption{Blob of CO emission to the SW of the star. The panels show of the $^{12}$CO $J=2-1$ emission (colour) and $^{13}$CO $J=2-1$ emission (white contours at $[3,5,10]\sigma$) across the velocity range $[17;26]$\,\kms. The extent of the colour scale has been limited for the sake of visibility. The stellar position is indicated as a black cross (+), the NOEMA beam is plotted in the bottom left corner as a white filled ellipse. }
    \label{fig:swblob}
\end{figure*}

\subsubsection{SW blob}\label{sect:swblob}
We find an emission blob in the $^{12}$CO and $^{13}$CO emission to the SW of the star extending from roughly 5\arcsec\/ to 6\farcs7 from the star, at $PA=245^{\circ}$ with respect to the stellar position, covering an opening angle of roughly $30^{\circ}$ . It is most apparent at $[17,26]$\,\kms, but likely already present at less red-shifted velocities (Figs.~\ref{fig:swblob} and \ref{fig:co_fullfov}). We note that the 29\,\kms channel bears no sign of this clump at all beyond two $3\sigma$ spots of only ${\approx}0\farcs3$ across.  

The SW blob is not measurably traced by any of the other emission lines in our data set and we find no spatially coincident continuum emission at our current sensitivity. We note that the $PA=245^{\circ}$ is similar to that of component A ($PA=225^{\circ}$), but we recall that the blue-shifted absorption of spectral line emission led us to derive a location in front of the plane of the sky for component A, whereas the SW blob in CO is found at significantly red-shifted velocities, suggesting a location behind the plane of the sky. 

Because we detected the blob in only one transition of $^{12}$CO and $^{13}$CO ($J=2-1$) and the emission is highly complex, we cannot reliably estimate the temperature or (CO) mass in the SW blob. We are not aware of any previously reported observations tracing this blob.

\begin{figure*}
    \subfigure[$PA=65^{\circ}$. $^{12}$CO (colour map), $^{13}$CO (white), SO$_2$ (black). \label{fig:pv_co_13co_so2_65}]{\includegraphics[width=0.5\linewidth]{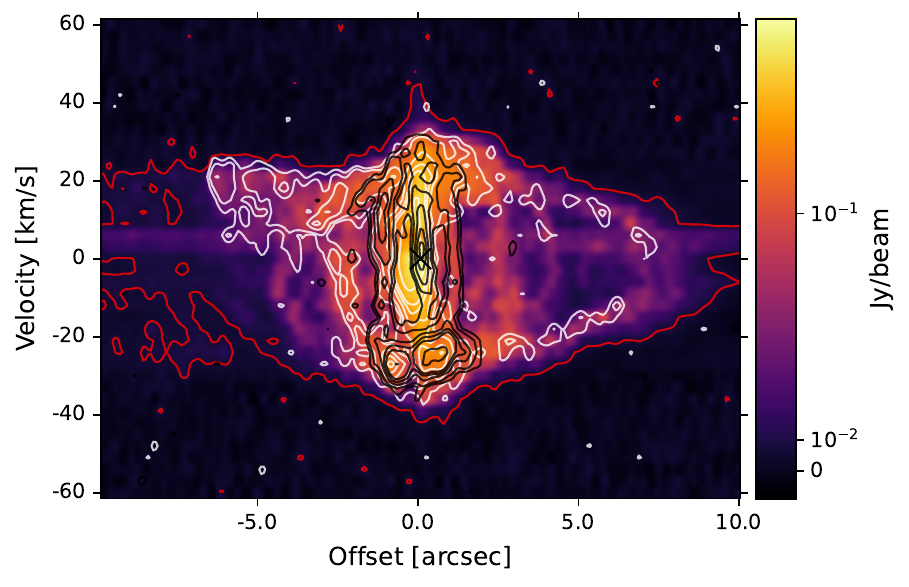}}
    \subfigure[$PA=65^{\circ}$. $^{12}$CO (colour map), SiO (white), H$_2$S (black).\label{fig:pv_co_sio_h2s_65}]{\includegraphics[width=0.5\linewidth]{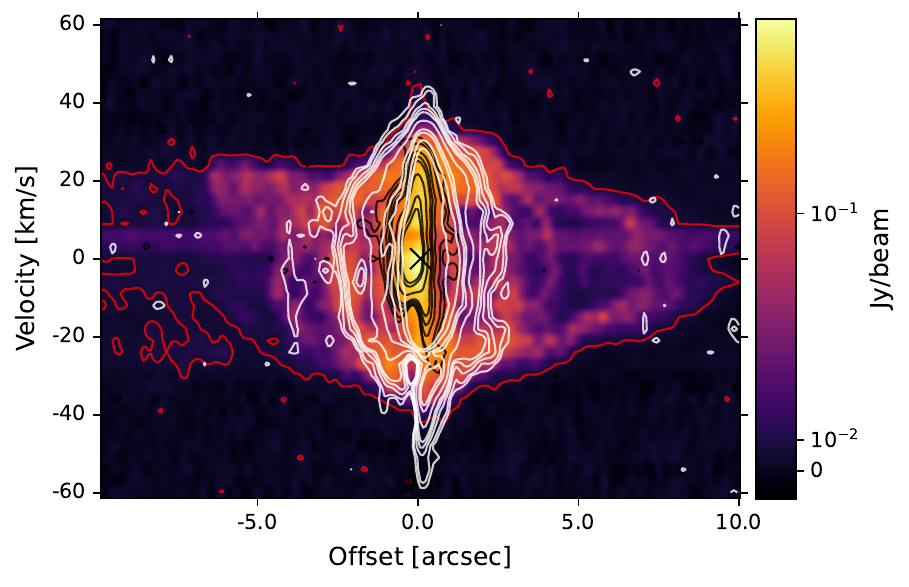}}
    \subfigure[$PA=155^{\circ}$. $^{12}$CO (colour map), $^{13}$CO (white), SO$_2$ (black). \label{fig:pv_co_13co_so2_155}]{\includegraphics[width=0.5\linewidth]{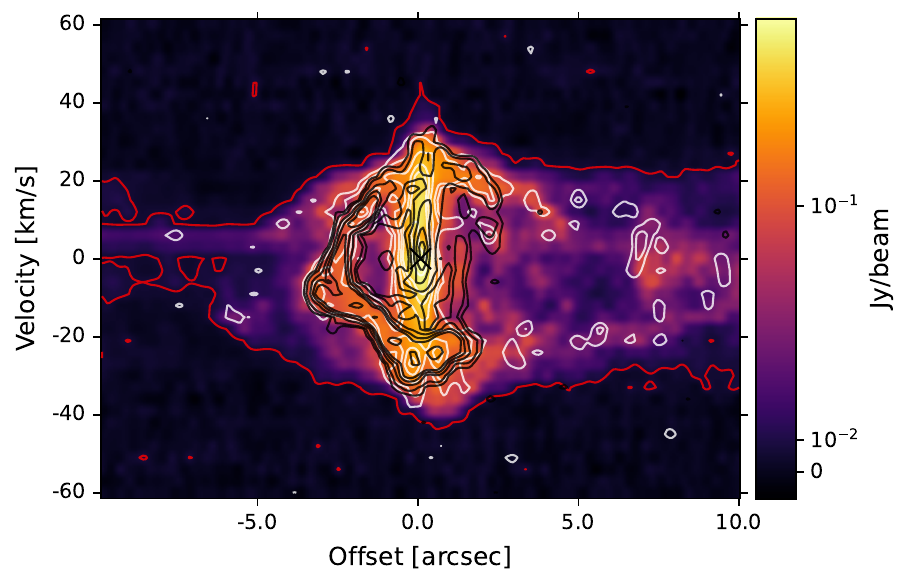}}
    \subfigure[$PA=155^{\circ}$. $^{12}$CO (colour map), SiO (white), H$_2$S (black).\label{fig:pv_co_sio_h2s_155}]{\includegraphics[width=0.5\linewidth]{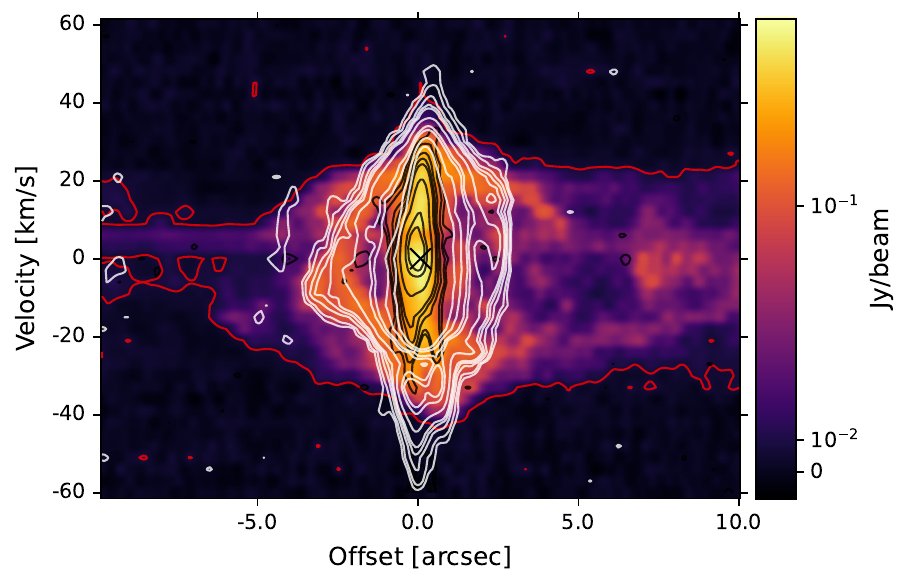}}
    \caption{Position-velocity diagrams. Cuts are centred on the assumed stellar position (marked as {\sffamily{X}}) and at position angles of $65^{\circ}$ to cut through the SW clump (top) and perpendicular to that, at $155^{\circ}$ (bottom). Contour levels are drawn at $[3,5,10,15,20,40,80,100]\times\sigma$ with $\sigma=2.0$\,mJy/beam; for $^{12}$CO, we only show the $3\sigma$ contour (red). Note: the CO emission at 5\,\kms and 8\,\kms is contaminated by ISM contributions.\label{fig:pvdiagrams}}
\end{figure*}

\begin{figure}
\centering
\includegraphics[width=\linewidth]{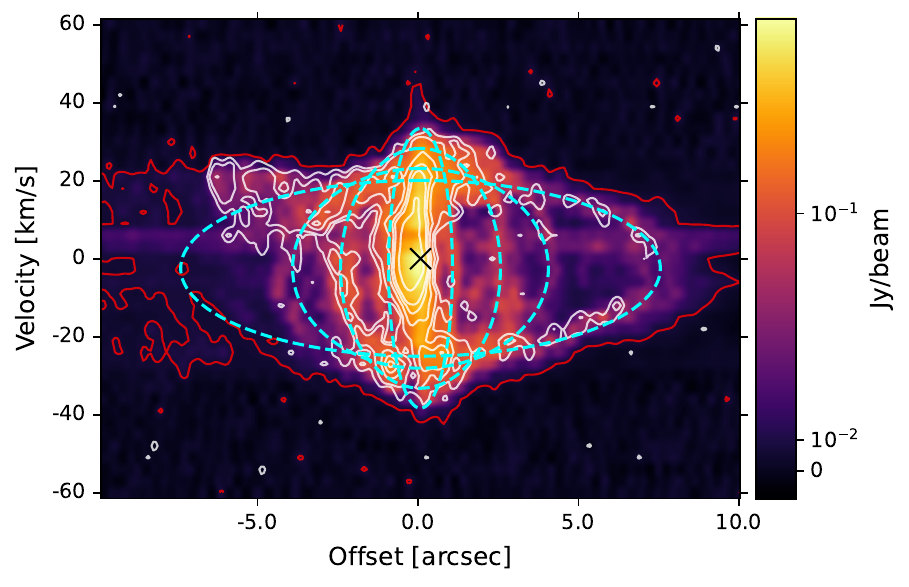}
\caption{ Same as Fig.~\protect{\ref{fig:pv_co_13co_so2_65}}, but omitting SO$_2$ contours and adding spherical shells centred on the star with radii $1\farcs0$, $2\farcs5$, $4\farcs0$, $7\farcs5$, and radially outward directed velocities of 35\,\kms, 30\,\kms, 25\,\kms, and 22\,\kms, respectively (dashed cyan lines). }
\label{fig:pv65_markup}
\end{figure}

\subsubsection{Arcs and shells}\label{sect:arcs}
We used position-velocity (PV) diagrams to highlight some of the substructures and (a)symmetry in the outflow. Figure~\ref{fig:pvdiagrams} shows PV-diagrams for cuts through the stellar position along $PA=65^{\circ}$, cutting also through the CO clump to the SW of the star,  as well as perpendicular to that, at $PA=155^{\circ}$, that is, roughly along the SE-NW axis identified in earlier maser and dust observations \citep{blocker_2001,boboltz_2000_siomasers_nmlcyg,etoka_diamond_2004,richards_1996_masers_h2o}. The cut along $PA=65^{\circ}$ and through the stellar position results in the best alignment in the PV-diagram of the CO emission with a central ellipse representing a spherical shell of ejected mass out to a radius of 1\arcsec\/ and at a velocity of $\approx35$\,\kms. Cuts along other position angles (through the stellar position) lead to worse aligned CO emission in the PV-diagrams, mostly noticeable as spatial offsets at the extreme negative velocities. This is suggestive of some deviation from spherical symmetry in the CSE already within about 1\arcsec\/ from the star, but our current observations do not provide the necessary angular resolution to characterise this structure in detail.

Figure~\ref{fig:pv_co_13co_so2_65} clearly shows the SW blob, extending (to the top left) at red-shifted velocities and negative spatial offsets. In addition to this, we see multiple arcs in the PV-diagrams of the CO emission, suggestive of (partial-) shell-like substructure in the CSE and reminiscent of structures seen in the outflows of several AGB stars \citep[e.g.][]{guelin2018_irc10216_shells,decin_2020,randall2020}. From a manual exploration of the PV-diagrams, we find that the contrast between the arc and inter-arc emission ranges from a factor of a few to almost ten. However, the complexities in the PV-diagrams are too many to get accurate contrast values. We propose the presence of multiple, likely incomplete shells of material having been ejected with possibly different velocities during episodes of enhanced mass loss. The tell-tale arcs are seen in the PV-diagram of CO for the cut at $PA=65^{\circ}$, where they appear reaching out to offsets of roughly $-4\farcs5$, $-2\farcs5$, $-1\farcs2$, $1\farcs0$, $2\farcs5$, $4\farcs0$, and $7\farcs5$. In Fig.~\ref{fig:pv65_markup}, we show what spherical shells of a given radius and moving at a given expansion velocity would look like in the PV-diagram. The largest of the arcs is seen only for positive offsets, that is, corresponding to a NE offset from the stellar position. 

The CO emission is too complex and the angular resolution too low to determine with certainty whether the shells are concentric. Similarly, it is difficult to constrain the velocities corresponding to the respective arcs with high precision, but a by-eye fitting in the PV-diagrams indicates that the largest shells are noticeably slower than the smaller ones, with $\approx$22\,\kms at 7\farcs5, $\approx$25\,\kms at 4\arcsec\/, $\approx$30\,\kms at $2\farcs5$, and $\approx$35\,\kms for $1\farcs5$. This significant decline in expansion velocity with increasing shell radius can be seen as a 'tightening' of the arcs in velocity space (Fig.~\ref{fig:pv65_markup}). This gradient in velocity could be a consequence of increasing wind velocities or deceleration of the material over time. The former could be linked to, for example, an increase in the stellar luminosity, the onset of a steadily increasing superwind, a change in the dust properties, or a change in the underlying ejection mechanism(s). On the other hand, it is not clear what would cause a deceleration of the CSE.

An additional complexity arises when we consider the PV-diagrams constructed along different position angles. In the PV-diagram for the cut at $PA=155^{\circ}$ (Figs.~\ref{fig:pv_co_13co_so2_155} and \ref{fig:pv_co_sio_h2s_155}), we found an arc extending out to roughly $10\arcsec$, but moving at the same velocity as the $7\farcs5$ shell at $PA=65^{\circ}$. A cut at an intermediate position angle leads to a PV-diagram, with an arc at an intermediate shell radius, but again with a similar velocity. This change in the size of the shell with position angle suggests that we are seeing ellipsoidal shells rather than spherical shells, with the major axis aligned roughly at $PA=155^{\circ}$. This is broadly in line with the previously identified dominant orientation of the infrared and maser emission \citep{blocker_2001,boboltz_2000_siomasers_nmlcyg,etoka_diamond_2004,richards_1996_masers_h2o}. A similar behaviour is also seen for the smaller shells. The uninterrupted structure in the PV-diagrams suggests that it is likely a continuous process that created these structures, but since the shell material seems to be moving at the same velocities (within the uncertainties of the by-eye fitting to the PV-diagram features), the ejection could not have taken place at the same time. It is therefore possible that we are seeing the imprint of a three-dimensional spiral, formed by mass being ejected while \nmlcyg moves on a significant orbit or by a binary companion affecting the outflow with an orbital plane along $PA=155^{\circ}$. Additional data at higher angular resolution are necessary to confirm this and obtain more information on a potential orbital configuration. 

A first estimate leads us to suggest a temporal spacing between the shells of the order of several hundred years, however, tightening by roughly a factor of two between consecutive detected arcs with estimated average ages of $\sim2800$\,yr, $\sim1400$\,yr, $\sim650$\,yr, and $\sim220$\,yr.

\begin{figure*}[htb]
\centering
\subfigure[OH 1612\,MHz and CO, red-shifted.]{\includegraphics[width=.3\linewidth]{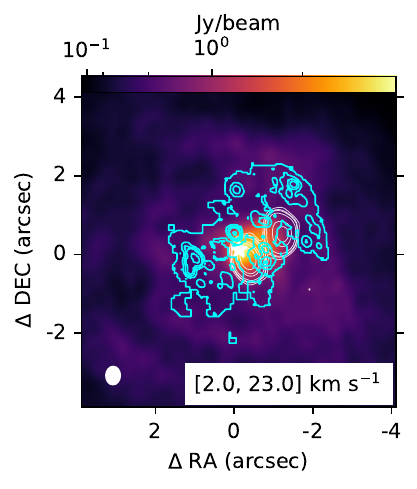}}
\hfill
\subfigure[OH 1612\,MHz and CO, blue-shifted]{\includegraphics[width=.3\linewidth]{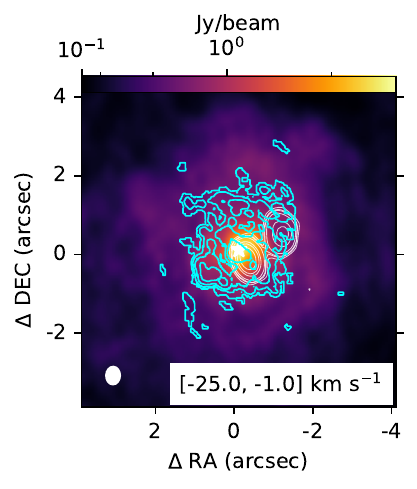}}
\hfill
\subfigure[OH 1665\,MHz and CO, blue-shifted.]{\includegraphics[width=.3\linewidth]{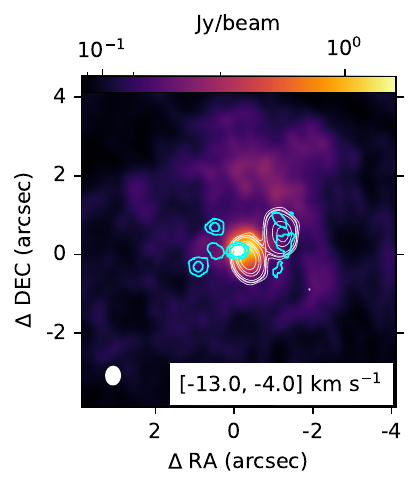}}
\hfill
\subfigure[OH 1612\,MHz and SO$_2$, red-shifted.]{\includegraphics[width=.3\linewidth]{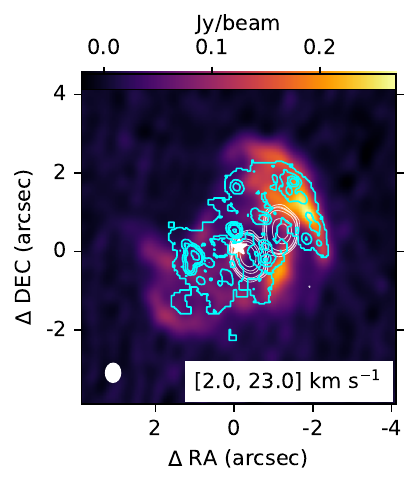}}
\hfill
\subfigure[OH 1612\,MHz and SO$_2$, blue-shifted .]{\includegraphics[width=.3\linewidth]{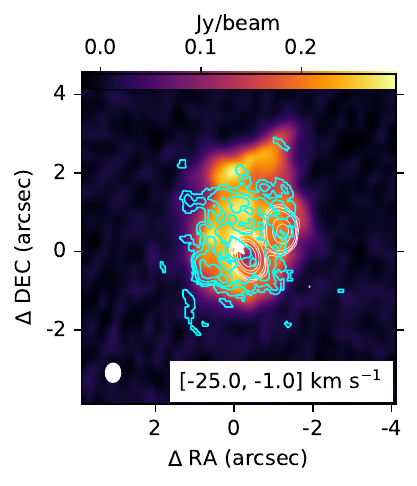}}
\hfill
\subfigure[OH 1665\,MHz and SO$_2$, blue-shifted. \label{fig:so2_oh1665_blue}]{\includegraphics[width=.3\linewidth]{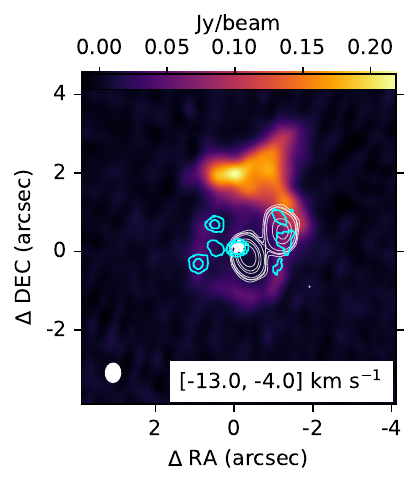}}
\caption{Comparison of OH maser emission and our NOEMA observations. The colour maps show moment-0 maps of CO $J=2-1$ (top) and SO$_2$ $J_{K_{\mathrm a},K_{\mathrm{c}}}=11_{1,11}-10_{0,10}$ (bottom) line emission (colour maps) constructed for the velocity ranges specified in each panel's lower right corner, in accordance with the ranges covered by the maser emission. Contours show the NOEMA continuum emission (white) and the OH maser emission from \citet{etoka_diamond_2004} in cyan. These are, from left to right: red-shifted OH 1612\,MHz, blue-shifted OH 1612\,MHz, and blue-shifted OH 1665\,MHz. The beams of the line emission measured with NOEMA are given in the bottom-left corner of each panel as white ellipses.
\label{fig:alignment_OH_lines_continuum}}
\end{figure*}

\subsection{Kinematics}\label{sect:kinematics}
The NOEMA observations and, in particular, the spectra and PV-diagrams as presented above, reveal significantly higher velocities in the wind of \nmlcyg than the $\approx$30\,\kms previously reported \citep{andrews2022, singh2021}. We detected CO emission across the velocity range $[-45;54]$\,\kms, with possibly some low-level emission even extending all the way to 65\,\kms. The SiO $v=0,J=5-4$ emission is seen across the range $[-62;55]$\,\kms, whereas the SiO $v=1,J=5-4$ emission is limited to the range $[-37;46]$\,\kms. We find SiS $v=0, J=12-11$ emission across $[-53,46]$\,\kms and SiS $v=1,J=12-11$ emission across $[-32,26]$\,\kms. In all of these cases, the highest-velocity emission originates from within one synthesised beam offsets from the stellar position. The 30\,\kms `threshold', however, is crossed at offsets of $\approx1\arcsec$ from the star for SiS $v=0$, at $\approx2\arcsec\/$ for SiO $v=0$, and for CO even at $\approx3\farcs5$. This would mean that the large-scale outflow is faster than previously assumed  and that there might be a very fast inner-wind component with speeds up to $\approx60$\,\kms. The latter would be in line with our earlier suggestion of a wind-speed gradient across the arcs traced in the PV-diagrams (Sect.~\ref{sect:arcs}) with increasing wind speed for decreasing arc radius and, hence, some intrinsic change in the wind parameters over time, possibly connected to a change in stellar parameters. However, we note that neither the H$_2$O nor the SiO maser observations were reported to show components at velocities beyond 30\,\kms \citep{richards_1996_masers_h2o,Zhang2012}, although both originate from the inner regions of the outflow, within one synthesised beam in the NOEMA observations. 

The high-velocity blue-shifted ($\leq-22$\,\kms) SiO $v=0$ emission is absent towards the SW portion of component A. Also, SiO $v=1$  seems to be absent in the directions of the SW part of component A and all of component B, but it is generally  not detected at equally large blue-shifted velocities ($\leq-40$\,\kms) either. We conclude (see also the discussion in Sect.~\ref{sect:molecular_continuum_emission}) that part of the high-velocity SiO is in front of and colder than the dust in component A. We deem the possibility that an optically thick component A obscures a compact high-velocity SiO cloud unlikely because (1) we detect emission in this direction at other velocities (and in other spectral lines) and (2) we would expect emission coming from small projected offsets from the peak of A to still enter the beam and be smeared out across the peak position and the SW portion of component A. However, this puts the high-velocity component at a significant spatial offset from the star, as the projected offset of the peak of component A is $\Delta\approx0\farcs3$ (Table~\ref{tab:cont_prop}), which corresponds to a lower limit to the distance to the star of $\approx4800$\,AU. Additional observations are needed to spatially resolve this inner region and better constrain both the dust and gas properties.

\section{Discussion}\label{sect:discussion}

\subsection{Comparison to maser emission}\label{sect:masercomp}
To get additional insight in the detailed structure of the CSE of \nmlcyg, we can compare the high-resolution maser emission reported in the literature to the continuum and line emission in our NOEMA observations. With our assumed stellar position (see Sect.~\ref{sect:stellarpos}), the  22\,GHz H$_2$O maser component to the NW of the star reported by \citet{richards_1996_masers_h2o} is located at the NW edge of component A (peak-$PA\approx-45^{\circ}$; projected offset $\approx290$\,mas; $v=5.5$\,km\,s$^{-1}$). All other 22\,GHz H$_2$O maser components reported by \citet{richards_1996_masers_h2o} and \citet{Zhang2012} are located so close to the stellar position that they coincide with the northern portion of component A and the region where we detect thermal H$_2$O and SiO emission. 

For lack of absolute positioning calibration, \citet{etoka_diamond_2004} assumed the stellar position to correspond to the brightest maser spot in the blue-shifted OH\,1612\,MHz maser emission. Considering the overall alignment of the OH\,1612\,MHz maser emission with the SO$_2$ emission (both red-shifted and blue-shifted; Fig.~\ref{fig:alignment_OH_lines_continuum}), this assumption appears to be valid. This positioning places one bright component of the blue-shifted OH\,1612\,MHz emission mostly concentrated at the NE tip of component A, at the stellar position, and one to the SE of the star. The blue-shifted OH\,1612\,MHz emission has no particularly bright or compact component at the projected position of component B. The blue-shifted OH\,1612\,MHz emission matches the blue-shifted SO$_2$ emission well in both extent and shape, with the exception of the northernmost SO$_2$ emission reaching beyond the OH emission. Similarly, the overall structure of the red-shifted OH\,1612\,MHz maser emission matches that of the red-shifted SO$_2$ emission well, with strikingly similar hook-like structures to the NW of component B. However, we find that the SO$_2$ emission extends in an arc-like shape beyond the OH maser emission to the SE of the star. We further also see an apparent avoidance in the red-shifted OH\,1612\,MHz emission of the peak of component B, which we previously discussed and suggested it is strongly correlated with red-shifted molecular emission. 

The blue-shifted OH\,1665\,MHz emission from \citet{etoka_diamond_2004} is distributed in compact components corresponding to the stellar position, three positions to the E and NE, and three weak, but significant components to the W, of which two overlap with the on-sky position of component B. The location of all but the stellar components is coincident with a ring-like structure seen in the SO$_2$ emission (Fig.~\ref{fig:so2_oh1665_blue}). 

Considering the strikingly similar offsets between our continuum components and the two strongest components in the blue-shifted OH\,1612\,MHz maser maps, we tested the viability of matching those pairs of emission, effectively putting the very brightest maser spot close to the peak of component B, rather than at the stellar position. Based on a worse agreement between the maser maps and several of the substructures identifiable in our CO and SO$_2$ maps, we are fairly certain that the original assumption is correct. This, in turn, means that we find some symmetry around the star between 1) continuum component B to the NW of the star (peak-$PA\approx-65^{\circ}$; projected offset $\approx985$\,mas), likely behind the plane of the sky (velocities of around 20\,km\,s$^{-1}$) and 2) the OH maser component SE of the star \citep[peak-$PA\approx 127^{\circ}$; projected offset $\approx940$\,mas;][]{etoka_diamond_2004} and in front of the plane of the sky (velocities of around $-20$\,km\,s$^{-1}$). Our observations hence provide further support for the finding of a receding NW lobe and an approaching SE lobe as traced by the H$_2$O masers on a smaller scale and the suggestion of a bipolar-like outflow embedded in a larger CSE \citep[e.g.][]{richards_1996_masers_h2o}.

\subsection{Mass-loss history} \label{sect:mdothistory}
The interpretation of the continuum components A and B as dust clumps implies directional, enhanced mass loss that has been sustained over significant timescales, combined with very efficient dust formation. Assuming a typical terminal velocity of 30\,km\,s$^{-1}$ and keeping in mind that we only measure a projected distance, such outflows could have occurred over timescales of 60 and 250 years for components A and B, respectively. However, as indicated in the previous section, the observations indicate a higher outflow velocity close to the star, making such estimates highly uncertain. At the same time, we note that the MIR emission presented by \citet{schuster2009} partially coincides with our component A and the `bridge' between our components A and B, hinting at a potentially more continuous process.

The timescales connected to convective cells, Alfv\'{e}n waves, and hot spots, where lifetimes of variation fall between 150 days and 3-6 years \citep{hartmann_avrett_alfven_1984,lopezariste_2018_convect_betelgeuse,Kervella2018,lopezariste2022}, are far shorter than would be required to explain the formation of the observed components A and B. A strong magnetic field, as observed for \vycma and \nmlcyg, could possibly support the formation of such large clumps \citep{ogorman2015,vlemmings2002_magfields,vlemmings2017_vycma}. Of course, higher-angular resolution observations might reveal these components to be made up of multiple smaller dust clumps and properties and formation timescales will need to be revised.

In contrast to these localised (limited opening angle) mass-loss enhancements, the likely presence of arcs in the outflow (see Sect.~\ref{sect:arcs}) is suggestive of repeated large-scale density or mass-loss enhancements, that is, covering a large part of the stellar surface. As mentioned, these might be a consequence of binary interaction, but based on the currently available data set, we are not able to constrain the properties of such a potential binary system. 

\subsection{Considering the origins of arcs and shells}\label{sect:arcs_origin}
In Sect.~\ref{sect:arcs} we estimated a declining temporal spacing between arcs, ranging from 2800\,yr to only 220\,yr. Although very long periods seem unlikely for a binary system containing an RSG \citep[e.g.][]{patrick2022_RSG_binaries}, the potential tightening of a binary orbit, ultimately resulting in a binary engulfment and a merger/common-envelope event is considered in theoretical models. Recently, \citet{landri2024_binaryRSGmassloss} explored the effect of a binary companion (2\,\msun) on an eccentric orbit around an RSG (20\,\msun), which grazes the RSG and finally plunges into it after a number of orbits. Although \vycma was considered a reference object for this study because of the high level of complexity in its CSE, observations of this object have  not revealed similar structures to those produced in the simulations thus far. The large-scale asymmetries of the CSE and the arcs and shells revealed by our observations of \nmlcyg, on the other hand, do exhibit reasonable similarity to the simulated density structures. On the other hand, we did not trace the same velocity structure as that simulated by \citet[][ their Fig.~8]{landri2024_binaryRSGmassloss}. This could be a consequence of the achieved angular resolution, preventing us from detecting more tightly wound arcs, or an indicator that this simulation is not an adequate representation of the \nmlcyg system. The separation between the detected arcs or shells of the order of a few hundred years is indeed considerably different from the $20-40$ year orbit of the simulated system and such a binary separation might induce different density and velocity structures. However, the tightening of the density enhancements we tentatively detect in our NOEMA observations are in line with a continuously shrinking orbit. We might hence be witnessing the inspiral of a binary companion into the RSG. The angular resolution of the observations ($\sim$0\farcs4) translates to a physical size scale of $\sim600$\,AU, or a timescale of $50-100$ yrs, for a $35-60$\,\kms outflow velocity. This means that there could be additional arcs in the CSE following the same tightening pattern as described above, but which we cannot spatially resolve.  Higher-angular resolution observations will be critical to test the plausibility of this ongoing binary inspiral scenario. 

As an alternative to the impact of a binary companion, we could consider the mass-loss enhancements as intrinsic to the star, similarly to detached shells in AGB stars \citep{olofsson1996,maercker2024_dcses}. These are brief mass-loss enhancements that arise as a consequence of thermal pulses. In these AGB stars, only one detached shell is ever seen at a time in CO, because of the long inter-pulse periods (of the order of $10^{4-5}$\,yr). In models for super-AGB stars, on the other hand, the inter-pulse periods go down to a few $10^2$\,yr \citep{siess2010_superAGB,doherty2014_superAGB}, with decreasing inter-pulse periods for an increasing initial stellar mass. However, the $25\,M_{\odot}$ initial mass estimate for \nmlcyg \citep{Zhang2012} is at odds with the super-AGB suggestion, given that the current framework of stellar models puts an upper limit to the initial mass of super-AGB stars to roughly $10\,M_{\odot}$ \citep{doherty2015}. Furthermore, the age of \nmlcyg for an assumed 25\,$M_{\odot}$ initial mass \citep[8\,Myr;][]{Zhang2012} is already significantly higher than the estimated age of the Cyg OB2 cluster ($2-3$\,Myr). This problem is only aggravated if the star had a much lower initial mass. In addition, the most massive super-AGB models by \citet[][for $9.8\,M_{\odot}$ initial mass]{doherty2015} reached a maximum luminosity of $1.11\times10^5\,L_{\odot}$, whereas \citet{Zhang2012} estimated $L_{\star}=2.7\times10^5\,L_{\odot}$ for \nmlcyg.  In conclusion, we deem it unlikely that \nmlcyg is a super-AGB star rather than an RSG if the distance and luminosity estimates by \citet{Zhang2012} are correct. Thus, we reject the characterisation of  the arcs as detached-shell analogues.

\subsection{Comparison to other RSGs -- \vycma in particular}\label{sect:otherRSGs}
Previous studies of RSGs have revealed complex outflows around \vycma, \mucep, Betelgeuse, and Antares \citep{Ohnaka_2014, Kervella2011, Kervella2018, kaminski_vycma_2019, Montarges2019, Humphreys_2022,Murakawa_2003,tabernero2021}. The similarities between the stellar properties of \vycma and \nmlcyg \citep{Zhang2012_vycma,Zhang2012} call for a direct comparison between their circumstellar environments, in particular. 

The amount of dust present around \nmlcyg appears to be similar to that around \vycma. Similarly to the case of \vycma, the continuum emission towards \nmlcyg is dominated by circumstellar contributions that reveal substructures asymmetrically offset from the star. \citet{ogorman2015} reported $4\times10^{-5}$\,\msun of hot ($\sim1000$\,K) dust close to the star and $2.5\times10^{-4}$\,\msun of warm ($<450$\,K) dust at a distance of at least 400\,AU from the star, based on observations at $\sim$325\,GHz and $\sim$659\,GHz. Adding measurements at $\sim$178\,GHz, \citet{vlemmings2017_vycma} reported a dust mass of at least $1.2\times10^{-3}$\,\msun with an upper limit to the dust temperature of $\sim$100\,K. \citet{humphreys2024_vycma_ALMA} reported ALMA detections at $\sim$250\,GHz of $\approx2.5\times10^{-4}$\,\msun of dust at temperatures in the range $300-500$\,K, which are not detectable in the optical or infrared. The dust-mass estimates we report for \nmlcyg suggest at least the same amount of dust close to the star, but possibly considerably larger masses in case of optically thick dust (see Table~\ref{tab:cont_prop}).

The extent of \nmlcyg's CO envelope also exceeds that of \vycma. Recent observations with ALMA of \vycma \citep{singh2023_vycma} have revealed an arc of emission of CO($2-1$) and HCN($3-2$) located at roughly 9\arcsec\/ offset NE from the star, considerably further out than the roughly 5\arcsec\/ traced by the CO($3-2$) emission presented by \citet{kaminski2013}, which might have been affected by the filtering out of some large-scale emission. As reported in Sect.~\ref{sect:extent}, we find a maximum CO extent for \nmlcyg of $\approx3\times10^{17}$\,cm in the SE direction. Taking into account the distance differences between \vycma and \nmlcyg, we traced CO emission out to distances up to a factor 1.7 larger than for \vycma. Similarly, the \nmlcyg CO emission-region size significantly exceeds the location of the furthest clumps of CO emission reported by \citet[][$\approx4\times 10^{16}$\,cm]{Montarges2019} for \mucep. 

Recently, high-angular resolution ALMA observations of \vycma revealed collimated \textit{Hubble}-type outflows accelerated to $\approx100$\,\kms, carving through inhomogeneous large-scale, slower ($30-40$\,\kms) ejecta \citep{quintanalacaci2023_vycma}. The current observations of \nmlcyg have not revealed equally collimated streams at such high velocities, but they do indicate that there are components in the circumstellar environment moving at higher velocities than the roughly 30\,\kms previously reported for this star \citep{singh2022_1mm,andrews2022}. Higher-angular resolution follow-up is needed to further localise and characterise these high-velocity components. We note that in Sect.~\ref{sect:kinematics}, we report a velocity gradient across the detected arcs around \nmlcyg, with the velocity increasing towards the star. We are not aware of similar results around other RSGs.

\subsection{Impact on SN explosion}\label{sect:SN}
A dense circumstellar medium surrounding the collapsing RSG is required to explain the observations of many SNe II \citep{morozova2018_SNII_denseCSM,yaron2017_SN2013fs_denseCSM}. Furthermore, asymmetries in the pre-SN circumstellar medium have a significant effect on the early detectability of the SNe via the probability of X-ray emission escaping the system and on the SN light curve. For example, \citet{singh2024_SN2023ixf_asymmetry, shrestha2024_SN2023ixf_polarimetry} reported the need for asymmetry in the dense circumstellar matter and clumpiness in the large-scale circumstellar environment of SN~2013ixf to explain the evolution of the polarisation of the SN emission in the first $\sim$80 days after core collapse. \citet{jencson2023} described the progenitor to SN~2013ixf as an RSG of $10^5$\,\lsun, with a pulsation period of $\approx1120$\,days and a pulsation amplitude of 0.6\,mag at 3.6\,$\mu$m, no bright pre-SN outbursts, $T_{\rm eff}=3500^{+800}_{-1400}$\,K, $M_{\rm init} = 17\pm4$\,\msun, and a pre-SN mass-loss rate of $3\times10^{-5}-3\times10^{-4}$\,\msunyr for an assumed wind velocity of 10\,\kms between 3 and 19\,yr before explosion. These stellar and wind properties are very close to those of \vycma and \nmlcyg \citep{Zhang2012_vycma,Zhang2012}.

The model presented by \citet{singh2024_SN2023ixf_asymmetry} includes a confined dense circumstellar component with an $r^{-2}$ density profile, created by a short-lived mass loss at a rate of $10^{-2}$\,\msunyr in the inner $5\times 10^{14}$\,cm. However, there is no observational evidence for pre-SN episodic bursts \citep[][and references therein]{singh2024_SN2023ixf_asymmetry}. At larger radii, out to $10^{16}$\,cm, the material can be modelled with an $r^{-3}$ density profile, ejected from the star at an average $10^{-4}$\,\msunyr. Even though we observe somewhat higher expansion velocities in the material surrounding \nmlcyg, with velocities possibly reaching up to $\sim$60\,\kms, the average gas mass-loss rate of a few $10^{-4}$\,\msunyr  \citep{andrews2022} is in line with this pre-SN model. In addition, we find strong asymmetries in the CSE of \nmlcyg, in the form of rather massive, localised dust clumps, which could eventually give rise to a polarisation behaviour that is similar to what has been observed for SN~2013ixf. 

We note that \citet{singh2024_SN2023ixf_asymmetry} mentioned the possible scenario of a binary component impacting the circumstellar morphology and the pre-SN mass stripping from the RSG. We suggest that we could be seeing early stages of this in the case of \nmlcyg, considering that the arcs reported and discussed in Sects.~\ref{sect:arcs} and \ref{sect:arcs_origin} are possibly a consequence of a binary companion interacting with the star and/or the stellar outflow.

\section{Conclusion}\label{sect:conclusion}
We present NOEMA observations of the red supergiant \nmlcyg around 230\,GHz ($\lambda=1.3$\,mm) at an angular resolution of $\approx0\farcs4$. Our observations reveal the presence of two bright continuum components that we derived to be the emission from warm dust close to, but significantly offset from the star. We estimated dust masses of roughly $10^{-3}\,M_{\odot}$ for both components and temperatures of the order of 440\,K and 265\,K. We find that the cooler component is likely to be located in the red-shifted outflow and, hence, behind the plane of the sky that cuts through the star -- and vice versa for the warmer component. 

We detected $\sim$70 spectral lines across the observed spectrum ($214.2-222.3$\,GHz and $229.7-237.8$\,GHz)  and present maps for some transitions of selected molecules, such as CO, SO$_2$, and H$_2$S. The molecular lines trace multiple components in the outflow, including blobs, arcs and shells and reveal higher-velocity components than previously identified for \nmlcyg. A comparison to previously published OH maser maps confirms our assumption with respect to the stellar position and highlights a strong SE-NW orientation in the system. We propose that some of the observed features could be linked to an interaction with a hitherto unknown binary companion located in an orbital plane oriented along that same SE-NW direction. 

It is likely that higher-angular resolution observations will reveal a more complex substructure of both the dust and gas in the circumstellar environment of this source and that masses, temperatures, and timescales will need to be revised. Observations at other wavelengths will help constrain the dust properties.

\begin{acknowledgements}
This work is based on observations carried out under project number w20bi001 with the IRAM NOEMA Interferometer as well as complementary single-dish observations carried out with the 30m telescope. IRAM is supported by INSU/CNRS (France), MPG (Germany) and IGN (Spain). We thank Ka Tat Wong for his work as the local contact astronomer and his help and advice in the reduction and calibration of the observations. EDB and HA acknowledge financial support by Chalmers Gender Initiative for Excellence (Genie). G.Q. would like to thank funding support from Spanish Ministerio de Ciencia, Innovaci\'on, y Universidades through grant PID2023-147545NB-I00.
\end{acknowledgements}

\bibliographystyle{aa} 
\bibliography{aa54748-25.bib}

\begin{appendix}
\onecolumn
\section{Supporting material}\label{sect:spectrum}

\begin{figure*}[h!]
\centering
\includegraphics[width=\linewidth]{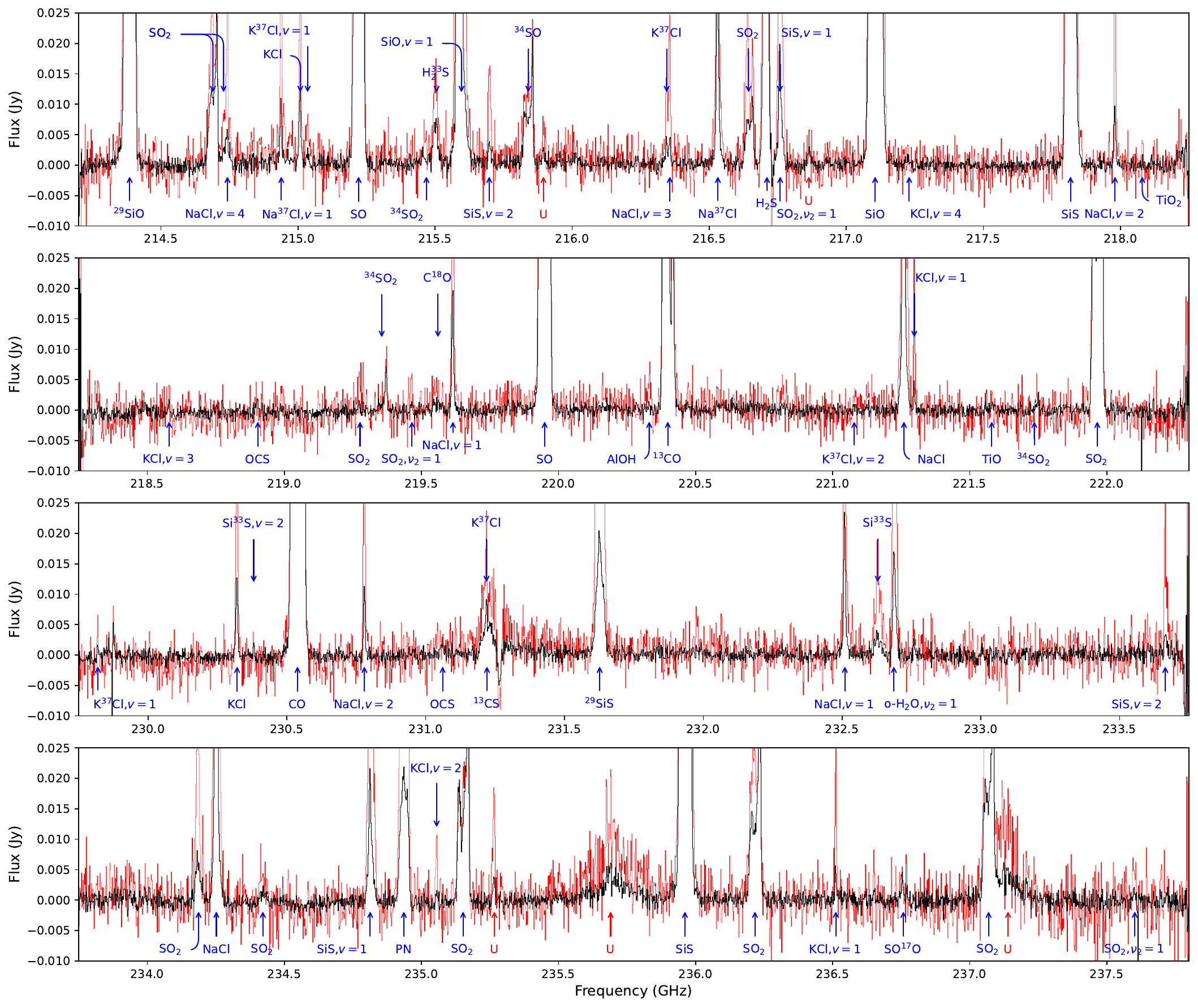}
\caption{NOEMA spectra of \nmlcyg: continuum-subtracted LSB (top panels) and USB (bottom panels) spectra extracted for the central beam (red) and for a 1\arcsec\/ aperture centred on the star (black). Molecular species responsible for the detected spectral lines are indicated in blue at the respective transitions' rest frequencies, unidentified features are marked in red as ``U''. }
    \label{fig:nmlcyg_noema_spectra}
\end{figure*}

Across the two covered frequency ranges $214.2-222.3$\,GHz (LSB) and $229.7-237.8$\,GHz (USB), we detect 67 spectral features from 15 species (26 species when counting isotopologues separately), of which about half are new detections compared to the observations presented by \citet{singh2021} and \citet{andrews2022}. Figure~\ref{fig:nmlcyg_noema_spectra} shows the spectra across the full frequency coverage towards the star and for a 1\arcsec\/ aperture centred on the star. Five features in these spectra could not be unambiguously identified with a carrier molecule and are marked as unidentified (U) in Fig.~\ref{fig:nmlcyg_noema_spectra}. 

\begin{figure*}
   \centering
   \includegraphics[width=\linewidth]{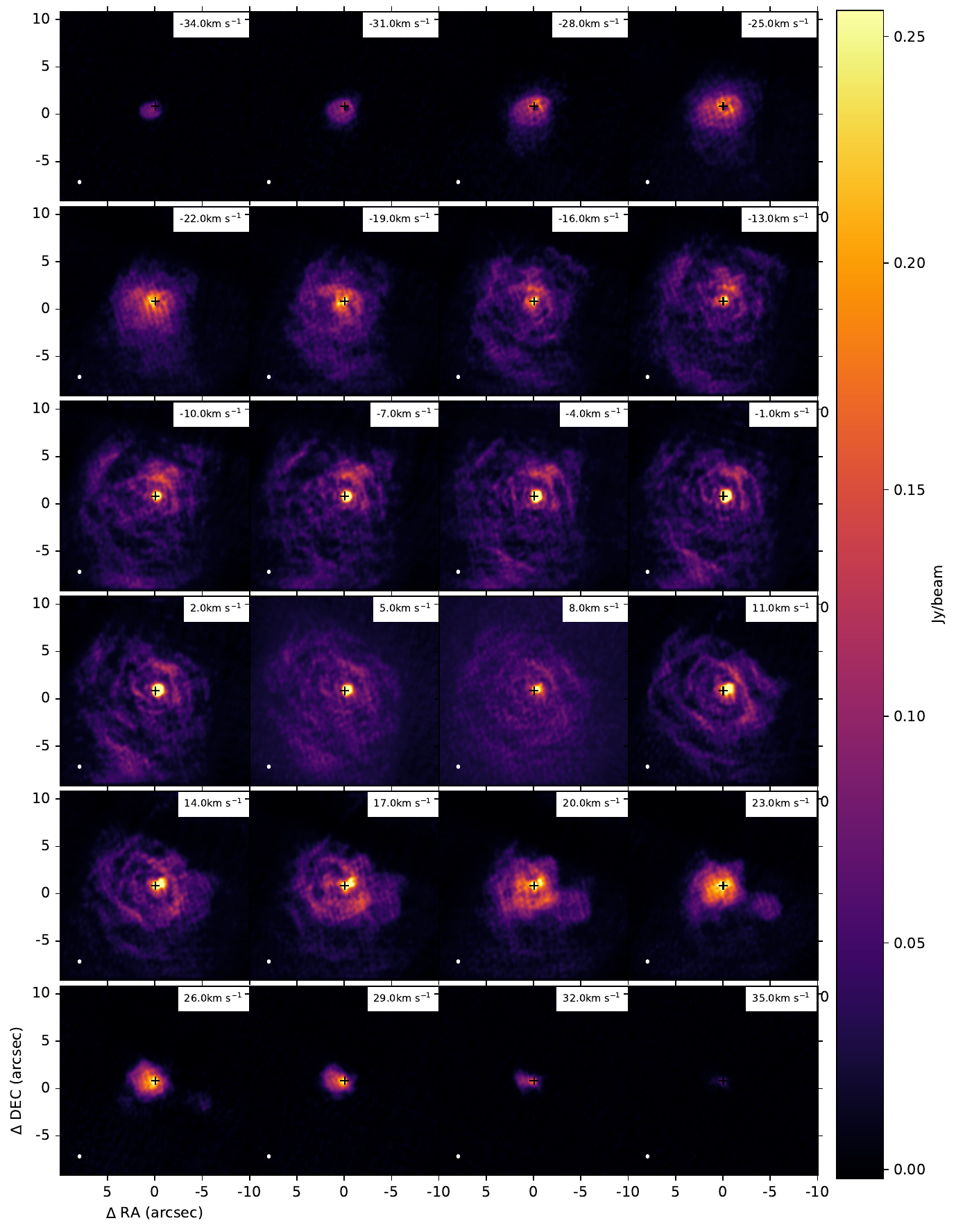}
   \caption{Channel maps showing  emission from CO $J = 2 - 1$ across the full field of view. The flux scale is cut off for the sake of visibility. Note: the channels at 5\,\kms and 8\,\kms are contaminated by ISM contributions spread across the entire field of view. The stellar position is indicated with a black {\bf +}.   }
   \label{fig:co_fullfov}
\end{figure*}

\begin{figure*}
    \includegraphics[width=\textwidth]{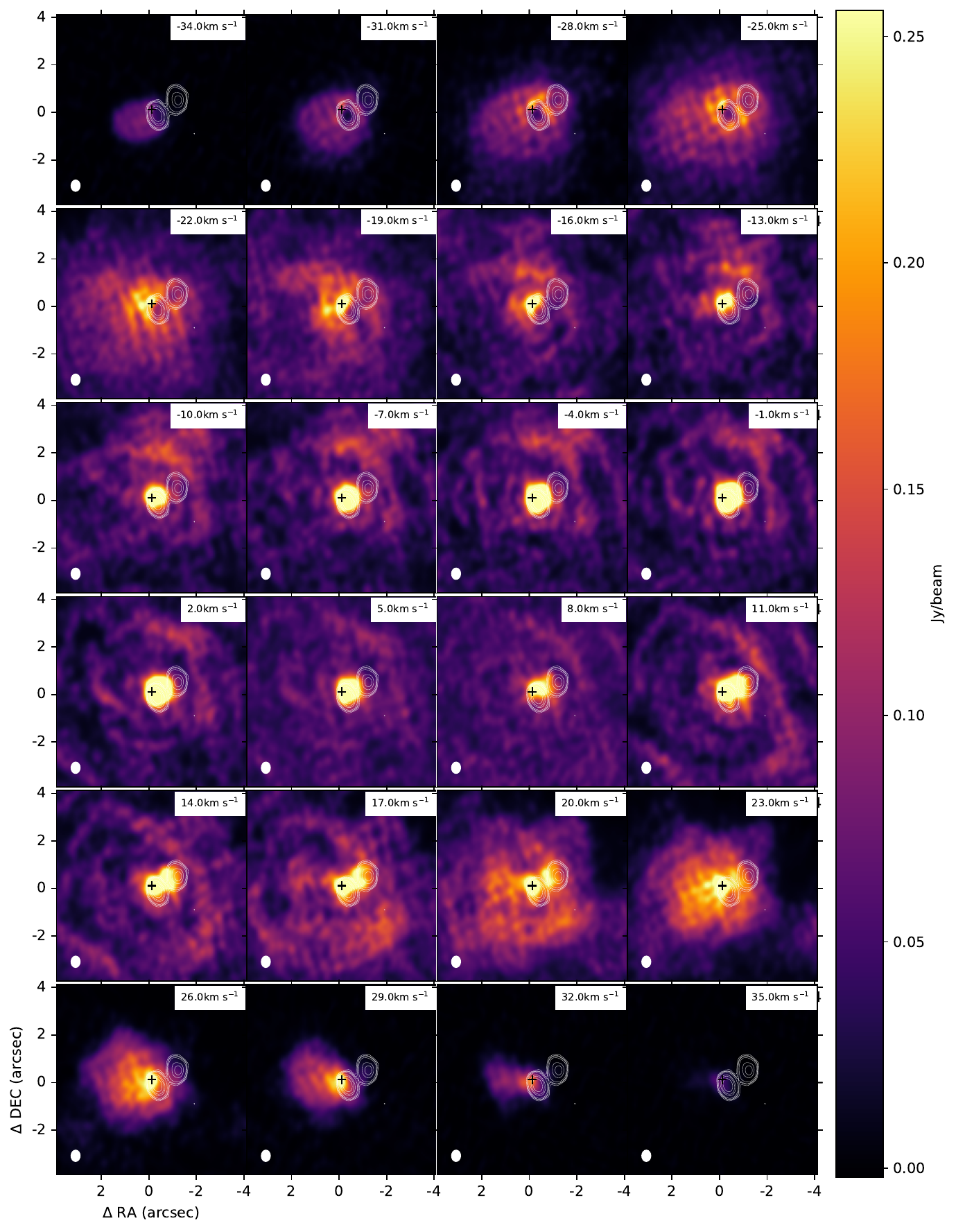}
        \caption{Channel maps showing  emission from $^{12}$CO $J = 2 - 1$ on a smaller scale than in Fig.~\ref{fig:co_fullfov}. The flux scale is cut off for the sake of visibility. Contours overlaid show the continuum emission (white) at  [5, 10, 20, 30, 40, 60, 80, 100]$\sigma$. The stellar position is indicated with a black {\bf +}.
        \label{fig:12co-channels}
        }
\end{figure*}

\begin{figure*}
    \includegraphics[width=\textwidth]{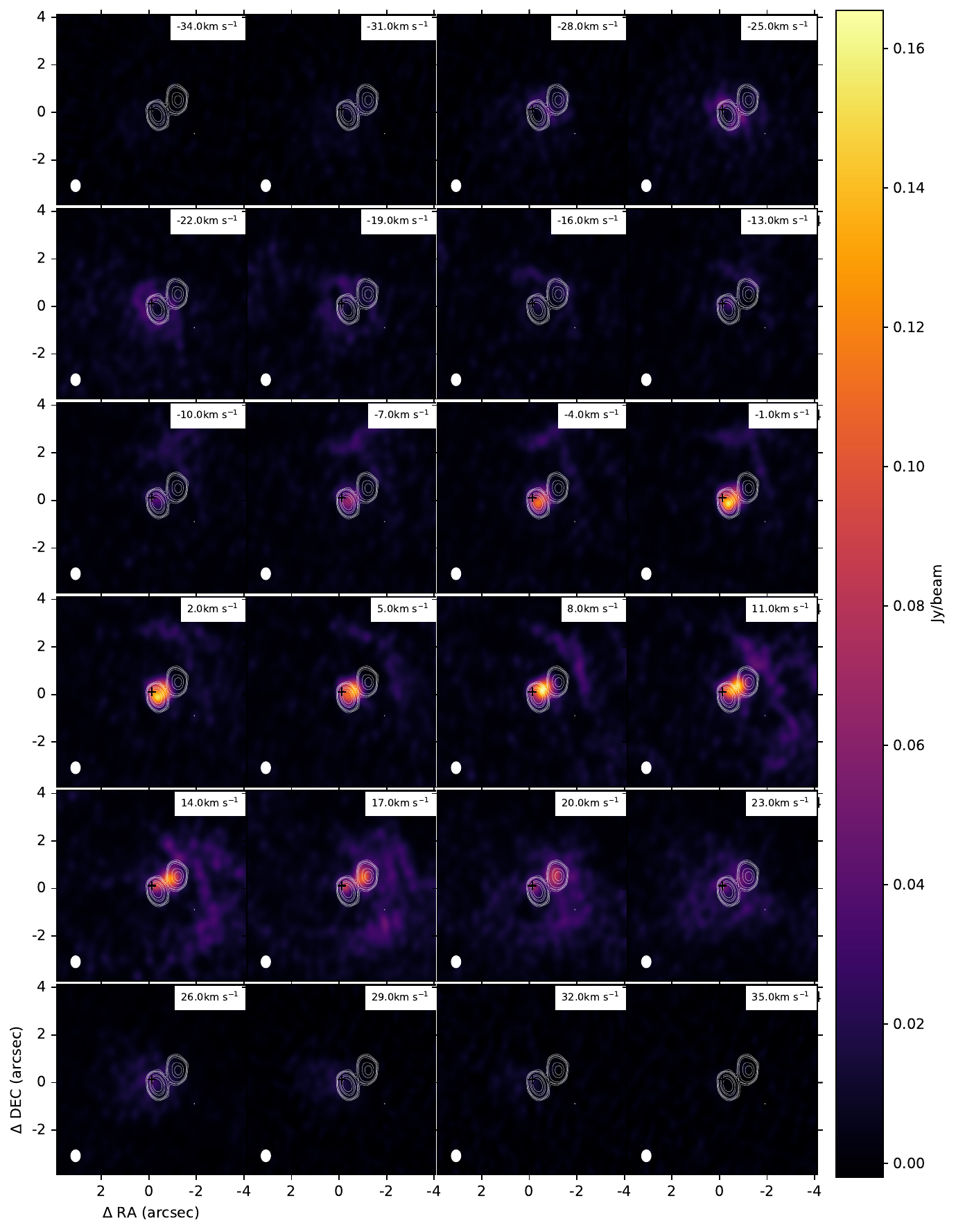}
        \caption{Channel maps showing  emission from $^{13}$CO $J = 2 - 1$. The flux scale is cut off for the sake of visibility. Contours overlaid show the continuum emission (white) at [3, 5, 10, 30, 50, 100]$\sigma$. The stellar position is indicated with a black {\bf +}.}
        \label{fig:13co-channels}
\end{figure*}

\begin{figure*}
    \includegraphics[width=\textwidth]{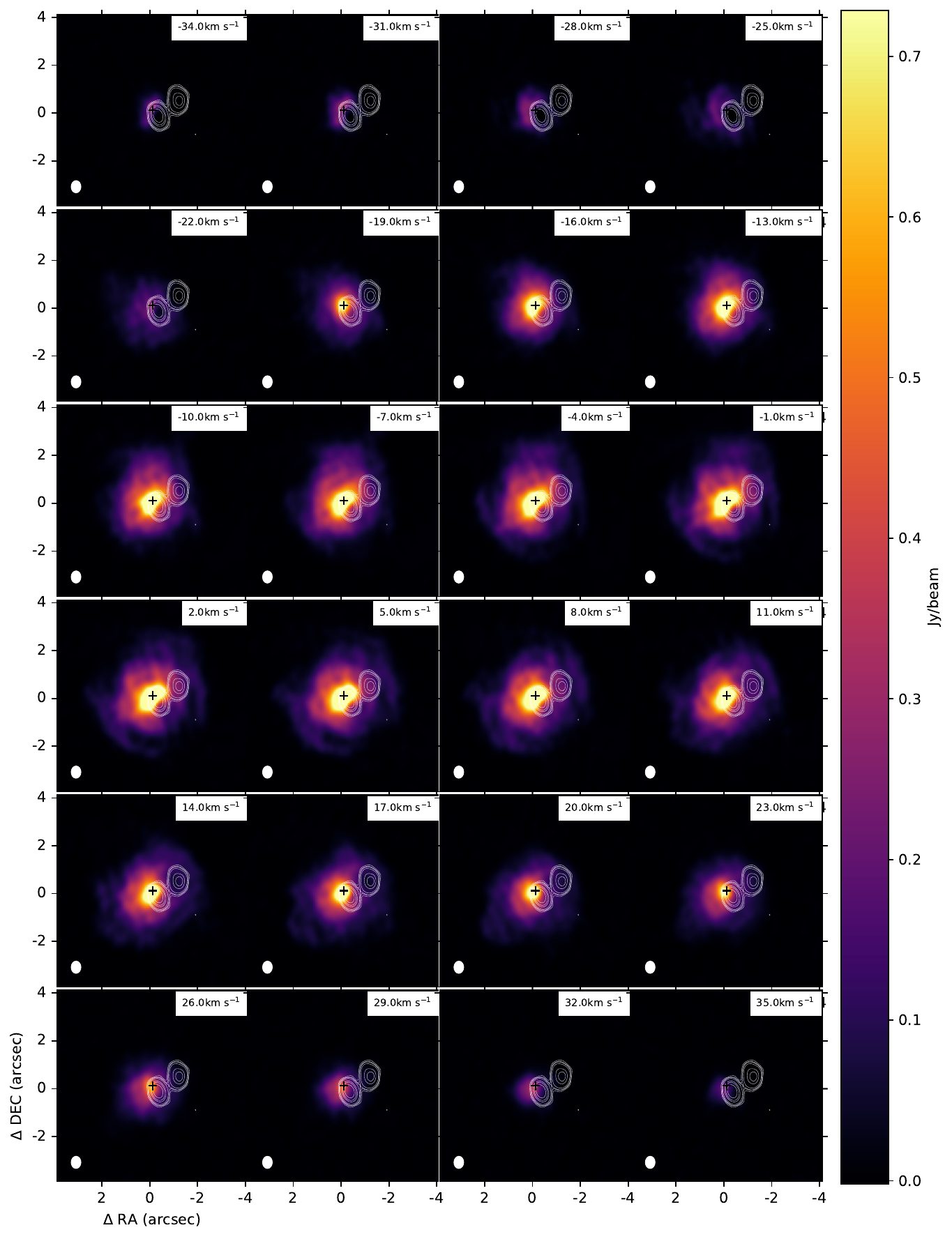}
        \caption{Channel maps showing emission from SiO $v=0,J=5-4$. The flux scale is cut off for the sake of visibility. Contours overlaid show the continuum emission (white) at [3, 5, 10, 30, 50, 100]$\sigma$. The stellar position is indicated with a black {\bf +}.
        \label{fig:SiOv0-channels} }
\end{figure*}

\begin{figure*}
    \includegraphics[width=\textwidth]{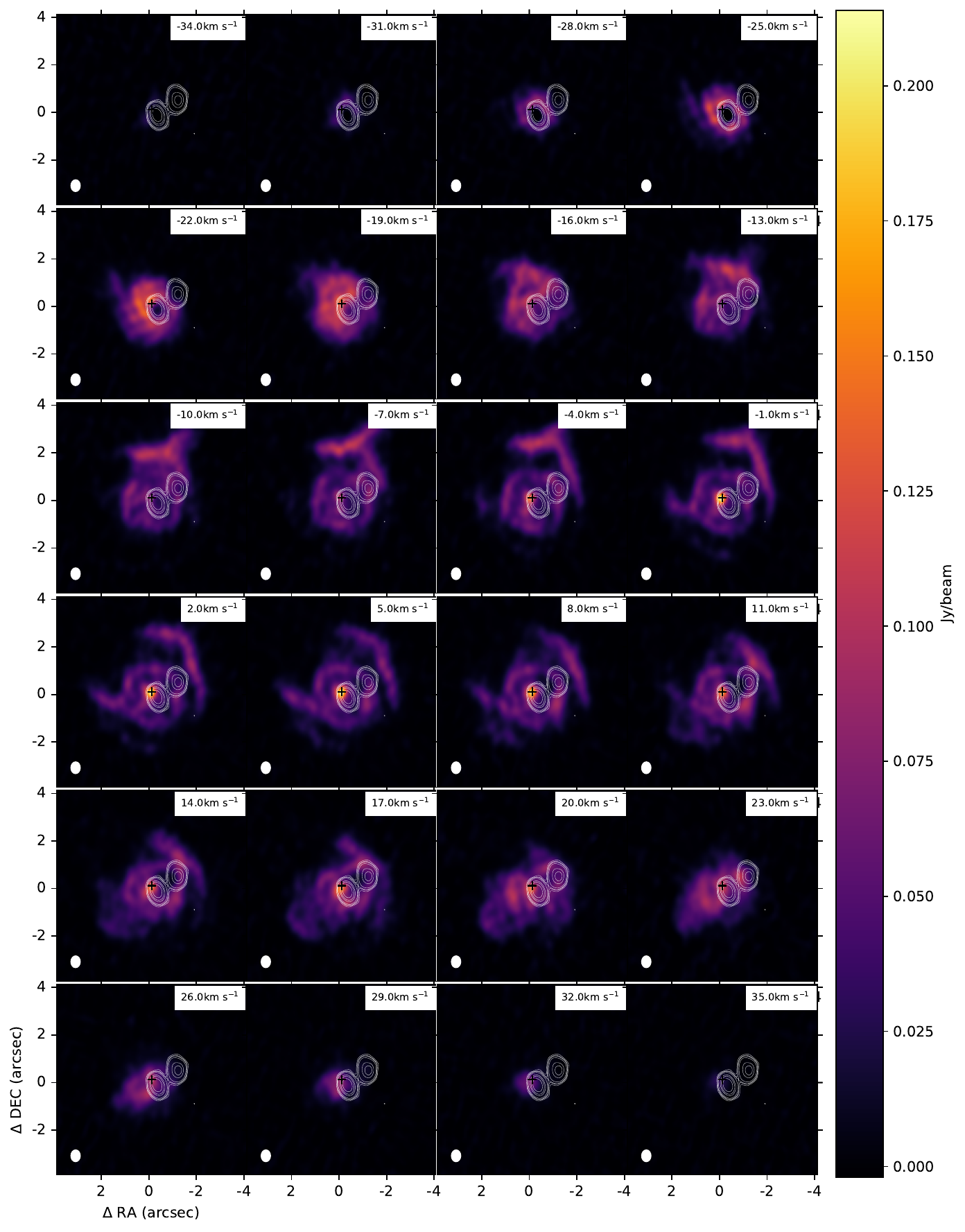}
        \caption{Channel maps showing emission from SO $J = 5_{6} - 4_{5}$. The flux scale is cut off for the sake of visibility. Contours overlaid show the continuum emission (white) at [3, 5, 10, 30, 50, 100]$\sigma$. The stellar position is indicated with a black {\bf +}.
        \label{fig:SO-5645-channels} }
\end{figure*}

\begin{figure*}
    \includegraphics[width=\textwidth]{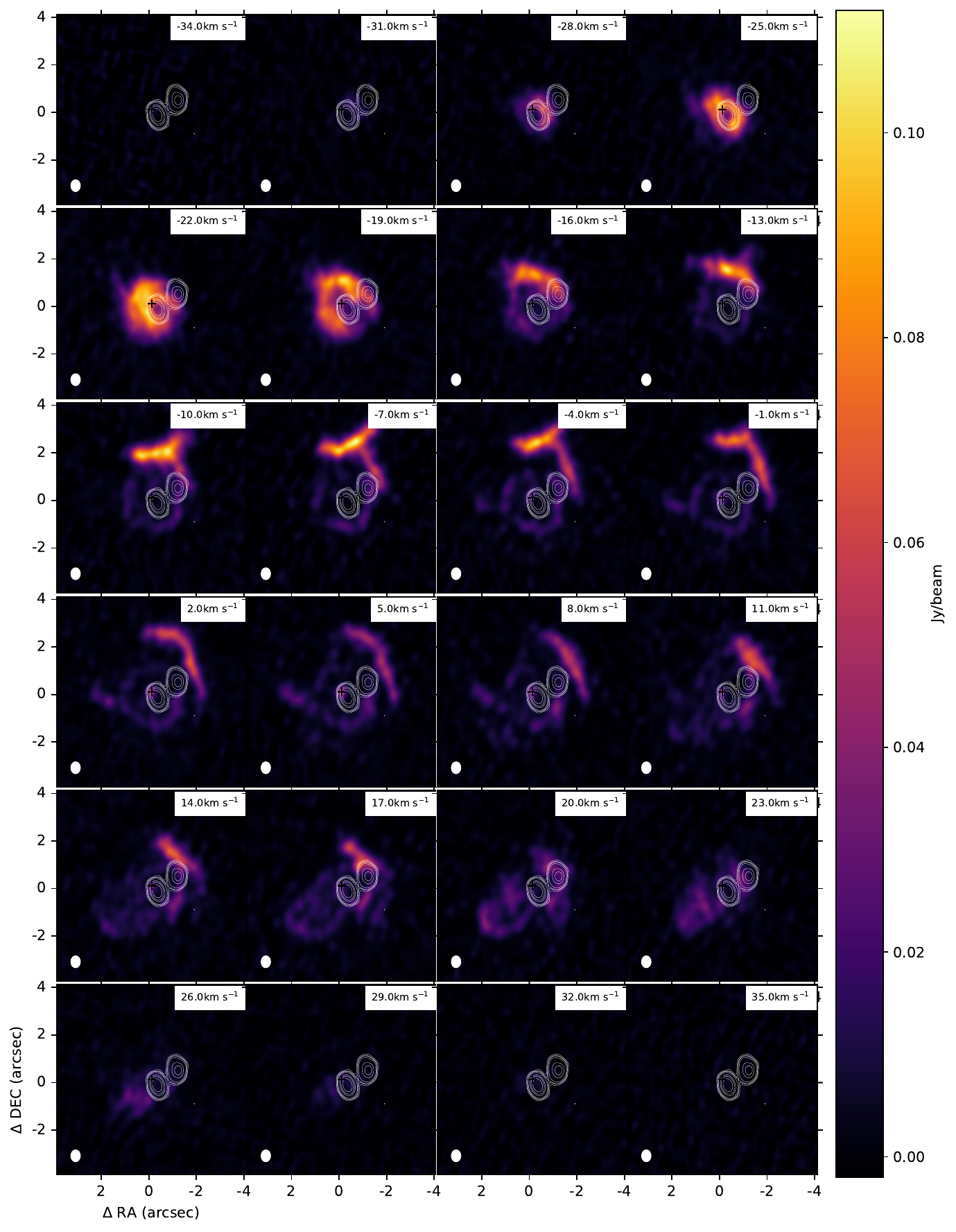}
        \caption{Channel maps showing emission from SO$_{2}$ $J = 11_{1,11} - 10_{0,10}$. The flux scale is cut off for the sake of visibility. Contours overlaid show the continuum emission (white) at [3, 5, 10, 30, 50, 100]$\sigma$. The stellar position is indicated with a black {\bf +}.
        \label{fig:so2_11-10-channels}
        }
\end{figure*}

\begin{figure*}
\includegraphics[width=\textwidth]{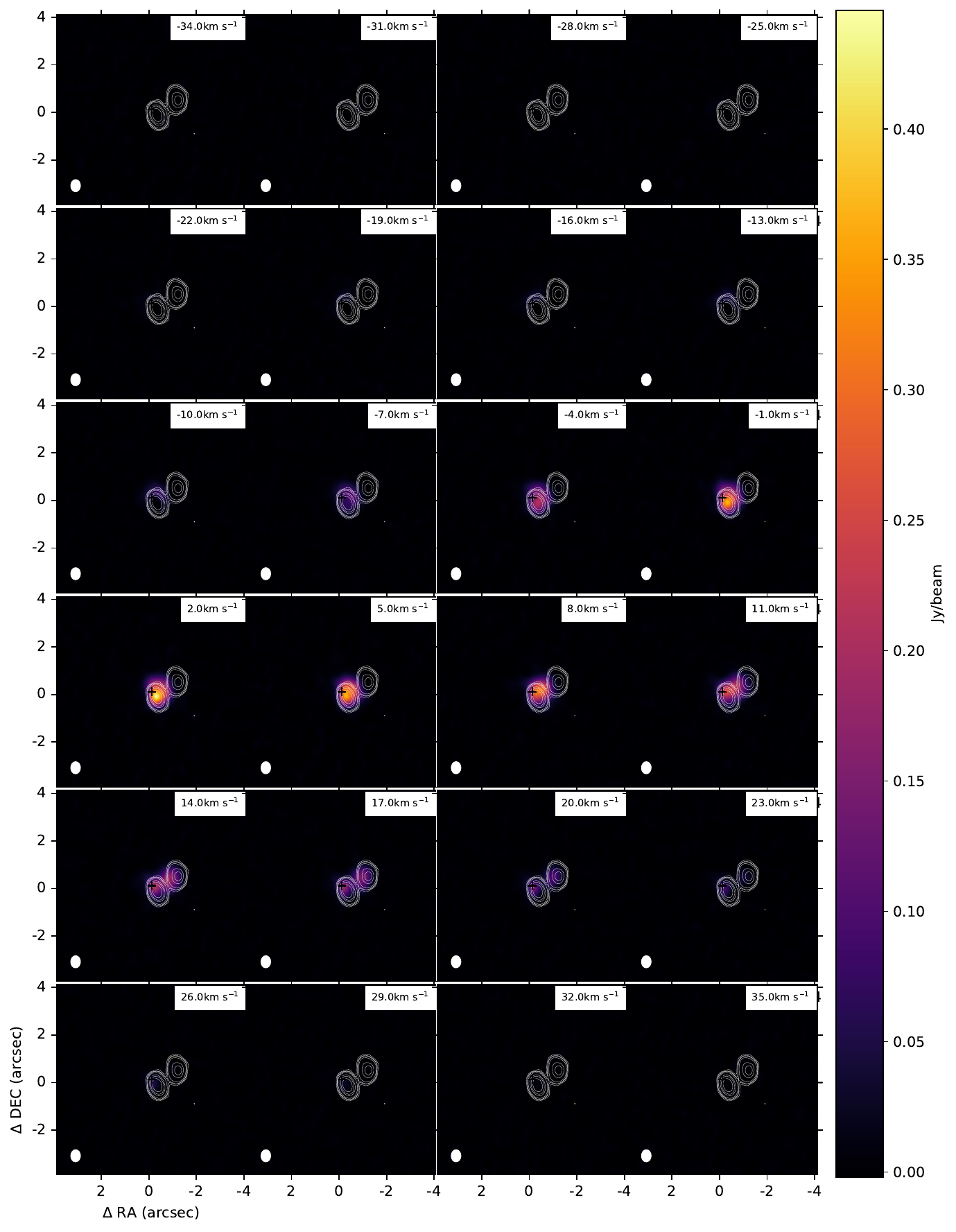}
\caption{Channel maps showing emission from H$_{2}$S $J = 2_{2,0} - 1_{1,1}$. The flux scale is cut off for the sake of visibility. Contours overlaid show the continuum emission (white) at [3, 5, 10, 30, 50, 100]$\sigma$. 
\label{fig:h2s-channels}}
\end{figure*}

\begin{figure*}[h]
\centering
\subfigure[SiO $v=1, J = 5-4$ \label{fig:siov1_mommaps}]{
\includegraphics[width=.9\textwidth]{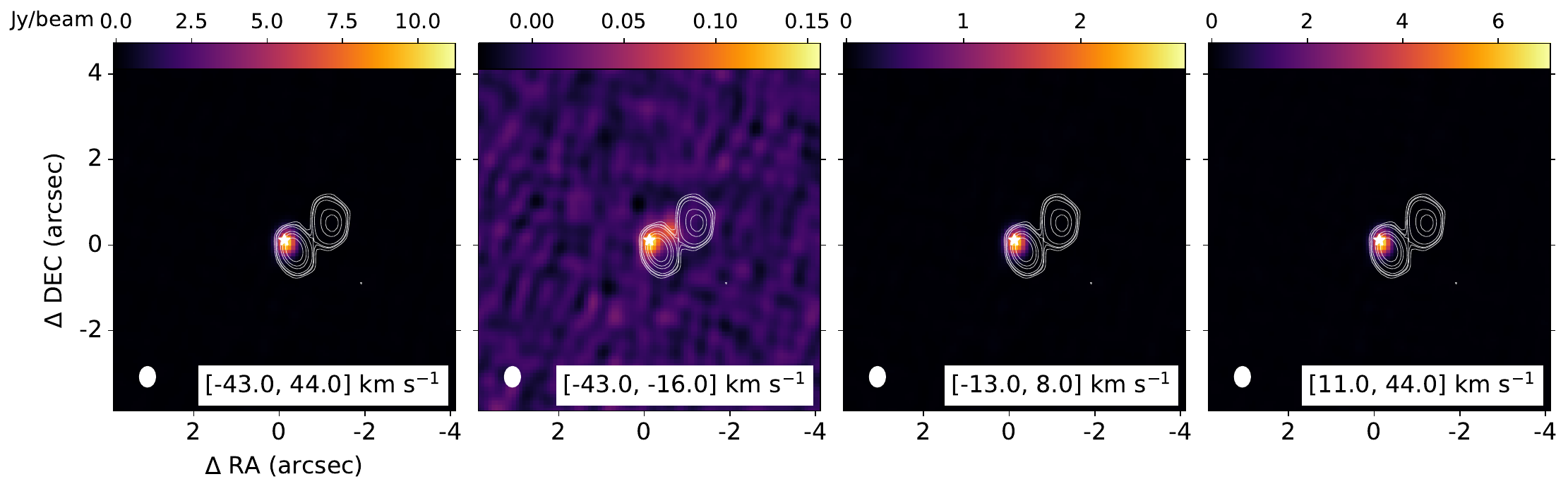}
}
\subfigure[SiS $v=0, J = 12-11$ \label{fig:sisv0_mommaps}]{
\includegraphics[width=.9\textwidth]{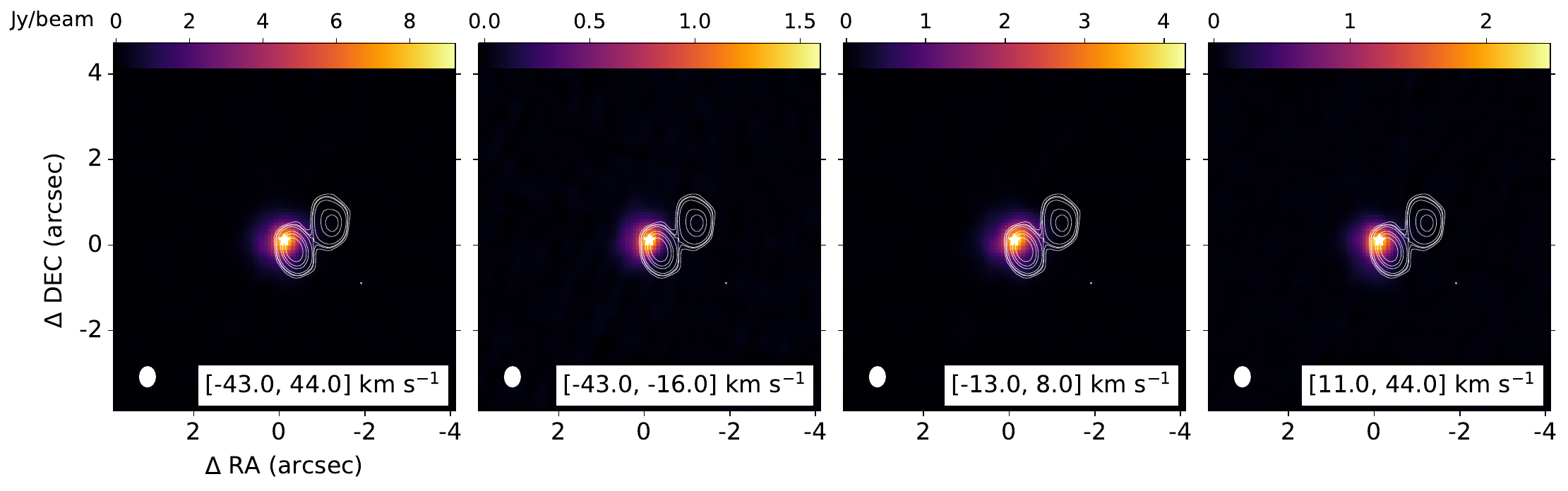}
}
\subfigure[SiS $v=1, J = 12-11$ \label{fig:sisv1_mommaps}]{
\includegraphics[width=.9\textwidth]{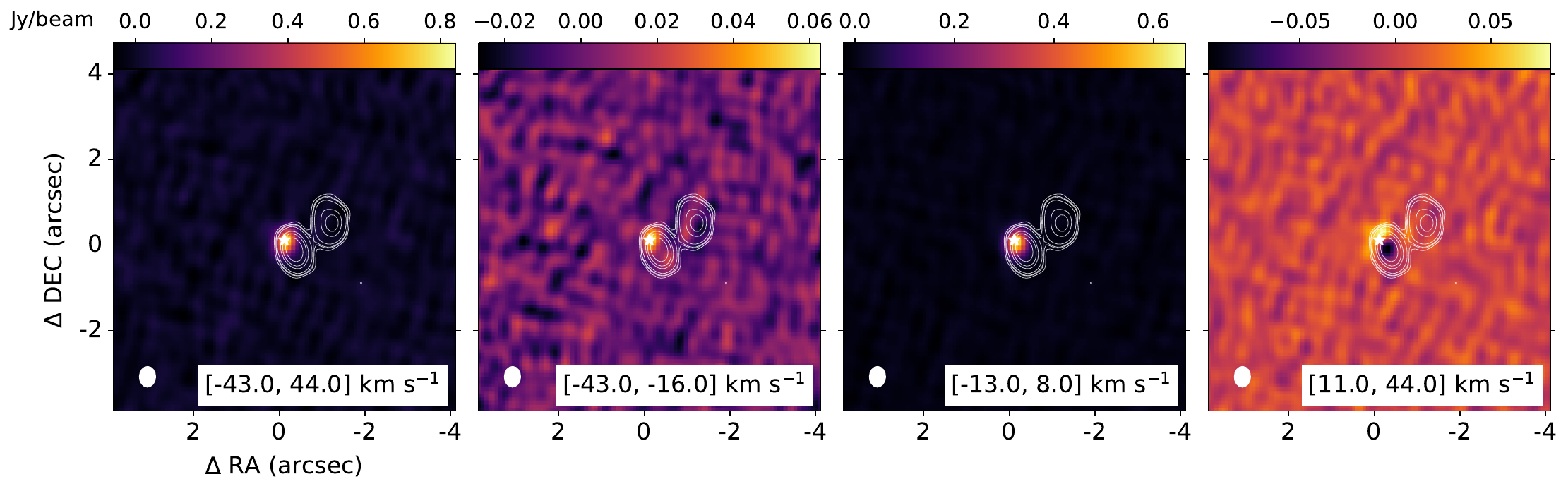}
}
\subfigure[H$_{2}$O $J = 2_{2,0} - 1_{1,1}$ \label{fig:h2o-mommaps}]{
\includegraphics[width=.9\textwidth]{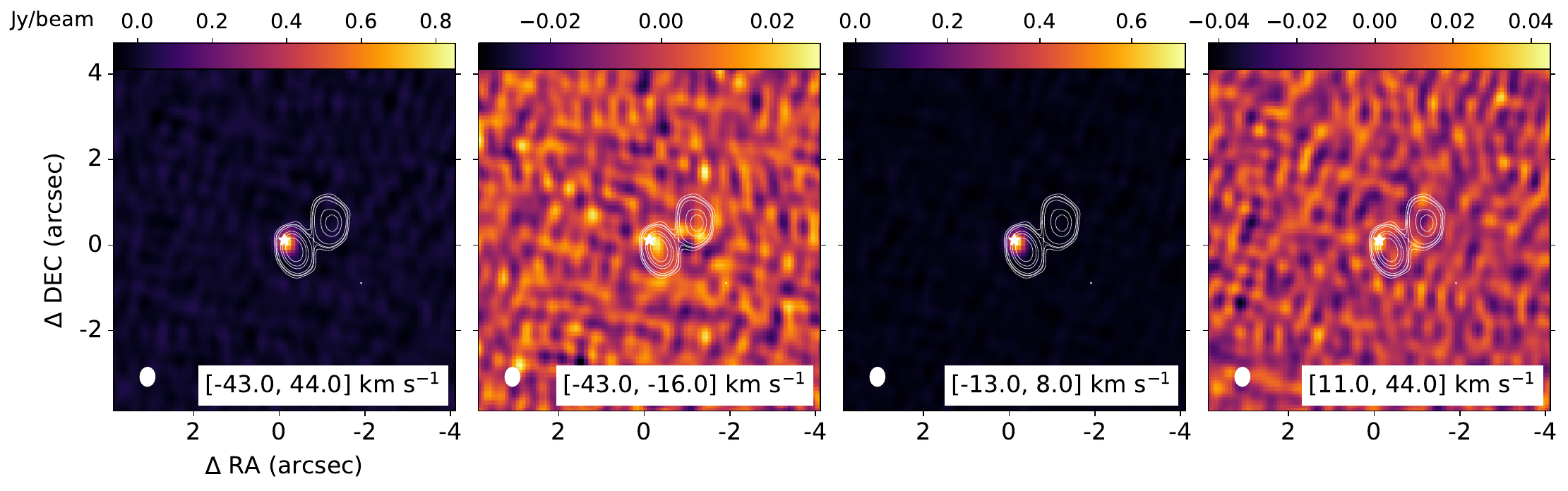}
}
\caption{Moment-0 maps for selected emission lines. From left to right:  overall moment-0 map,  blue-shifted, central, and red-shifted emission, covering the velocity ranges indicated in the bottom right corner of each panel. The stellar position is overlaid as a white star, and continuum emission plotted as white contours at [3, 5, 10, 30, 50, 100]$\sigma$.
          \label{fig:mom0_vranges_selection} }
\end{figure*}

Figures~\ref{fig:co_fullfov} -- \ref{fig:h2s-channels} show channel maps of selected emission lines of $^{12}$CO, $^{13}$CO, SiO, SO, SO$_2$, and H$_2$S and Fig.~\ref{fig:mom0_vranges_selection} shows moment-0 maps of selected emission lines of SiO, SiS, and H$_2$O that support the discussions in the paper. 

\vspace*{1cm}

\end{appendix}

\end{document}